\documentclass{jfm}
\usepackage{caption}
\usepackage{subcaption}
\usepackage{graphicx}
\usepackage{epstopdf, epsfig}
\usepackage{bbold}
\usepackage{color}
\usepackage{siunitx}
\usepackage{cmll}
\usepackage[mathcal]{eucal}
\usepackage{ucs}
\usepackage[utf8x]{inputenc}
\usepackage{amsmath}

\def\rr#1{(\ref{#1})}

\newcommand{\be}{\begin{equation}}
\newcommand{\ee}{\end{equation}}

\graphicspath{ {./Production_files_inertial_wavesI} }

\shorttitle{Traveling wave electroosmosis}
\shortauthor{A. Shrestha, E. Kirkinis and M. Olvera de la Cruz
}
\title{Universal behavior in traveling wave electroosmosis}
\author{A. Shrestha\aff{1,2}, E. Kirkinis\aff{1,3}
\corresp{\email{kirkinis@northwestern.edu} }
\and M. Olvera de la Cruz \aff{1,2,3}}
\affiliation{\aff{1}Center for Computation and Theory of Soft Materials, Northwestern University, Evanston IL 60208 USA \aff{2}Department of Physics and Astronomy, Northwestern University, Evanston, IL 60208 USA\aff{3} Department of Materials Science \& Engineering, Robert R. McCormick School of Engineering and Applied Science, Northwestern University, Evanston IL 60208 USA
}

\begin{document}

\maketitle

\begin{abstract}
Traveling wave charges lying on the insulating walls of an electrolyte-filled capillary, 
give rise to oscillatory modes which vanish when averaged over the period of oscillation. They also give rise to a \emph{zero mode} (a unidirectional, time-independent velocity component) which does not vanish. The latter is a nonlinear effect caused by 
continuous symmetry-breaking due to the quadratic nonlinearity associated with the electric body force in the time-dependent Stokes equations. In this paper we provide a unified view of the effects arising in traveling wave 
electroosmosis and establish the universal behavior exhibited by the observables. %of the system exhibit universal behavior.  
We show that the incipient velocity profiles are self-similar implying that those obtained with a single experimental configuration,
can be employed again to attain further insights without the need of repeating the experiment. Certain results from 
the literature are recovered as special cases of our formulation and we resolve certain paradoxes having appeared in the past. 
We present simple theoretical expressions, depending on a single fit parameter, that reproduce these profiles,
which could thus provide a rapid test of consistency between our
theory and future experiment. 
The effect becomes more pronounced when reducing the transverse 
dimension of the system, relative to the velocity direction, and increasing the excitation wavelength, and can therefore be employed for unidirectional transport
of electrolytes in thin and long capillaries. 
%We recover results from the literature as special cases of our formulation
%and resolve certain paradoxes that appeared before. }
General relations, expressing the zero mode velocity in terms of the electric potential and the geometry of the system only, can thus be easily adopted to suit alternative experimental settings. 
%This nonlinear effect is caused by a traveling wave electric field acting on a 
%traveling wave charge distribution, leading to, among others, a time-independent body force acting on the fluid.     
\end{abstract}

%--------------------------------------------------------------------------
% Maple calculations are in 
%---------------------------------------------------------------------------

% According to Eqs. (2.4.2) \emph{et seq.}, the pressure... Morse p. 1182
\section{Introduction}

Microfluidic devices can be classified according to their degree of configurability \citep[Box 2]{Paratore2022}. In the static state, 
the geometry of the device is set once and for all at the manufacturing level. In the configurable state, the user can format the 
device only to specific predefined states. And, in the \emph{reconfigurable} state the device can be rearranged at will in real time. As discussed by
\citet{Paratore2022}, traveling wave electroosmosis can be considered to belong to the third class (reconfigurable)
by shaping complex flow patterns using a suitable array of electrodes. Among its applications, this concept can be employed as a diagnostic
personalized tool employing \emph{very} low voltages ($1.5$ V, as was also predicted by \cite{Cahill2004}) and tunable frequencies, with low power consumption and 
cost-effective manufacturing based on recent metal-oxide-semiconductor technologies \citep{Yen2019}.

%When an electrolyte solution comes in contact with an insulating wall, a charge layer forms adjacent to the wall that brings the electrolyte into
%equilibrium. Mathematically, this is a boundary-layer problem for Gauss equation, whereby the highest order derivative is multiplied
%by a very small parameter (the Debye length, which characterizes the size of the charge layer). If in addition one applies an external electric field $\mathbf{E}$ parallel to the wall, 
%a body force $\rho \mathbf{E}$ is then exerted on the electrolyte with bulk charge distribution $\rho$. The charges next to the wall will be set into motion and since the electrolyte is viscous, it will drag the electrolyte along with it
%with a velocity $-\epsilon E\zeta/\eta$, where $\epsilon$ is the dielectric constant, $E$ is the amplitude of the applied electric field, $\eta$  is the dynamic viscosity of the electrolyte, and $\zeta$ is the value of the equilibrium potential at the wall \citep[ \S 6.5]{Probstein1994}. 
The physical effect of electroosmosis
employed in experiments for mixing and solute transport \citep{Stroock2000} has led to new advances, for instance, 
\citet{Zhao2020} showed that it can be employed in force measurements of electrolyte solutions by 
atomic force microscopy. New behavior includes the attraction of field lines towards
a surface charge discontinuity and commensurate macroscopic effects \citep{Yariv2004,Khair2008}. Moreover, under the application of a external field, induced-charge electro-osmotic flow has been found to develop over polarizable surfaces \citep{Squires2004}.

Traveling wave electroosmosis is a special, \emph{boundary-guided} electrokinetic effect whereby a body force is exerted on the bulk charges of an electrolyte
due to a traveling wave electric field (or voltage) excitation on the capillary walls. Since the liquid is viscous, it 
is being dragged along with the charges, thus providing control over its movement.  
Electrolyte pumping with traveling wave wall charges was introduced in the seminal, but largely overlooked, work of \citet{ehrlich1982} who showed 
numerically that an electrolyte separated by an electrode array with a thick dielectric layer, will move unidirectionally parallel to the charged wall - a velocity profile that will be termed \emph{zero mode} in the present paper.   
%for small field amplitude or frequency, and 
%also noticed backward pumping in the large parameter regime. 
\citet{Cahill2004,Ramos2005, Garcia2006} established the existence of this 
unidirectional flow in traveling wave electroosmosis experimentally
%thus partially confirming the 
%numerical results of \citet{ehrlich1982}. 
and proposed some theoretical models. Note that AC electroosmosis is a related effect whereby
AC signals in the kHz range are employed to generate regions of circulating liquid flow,
periodically changing their sense of circulation
\citep{Green2000,Gonzalez2000}. Invoking a geometrical asymmetry one can also generate unidirectional flow
cf. \citet{Ajdari2000}. 

%In the former paper it was found that the experimentally determined velocity was 
%significantly lower compared to its theoretical counterpart (same conclusions as those reached by \citet{ehrlich1982}), and in the latter, theory gave rise to zero velocity averaged over time, in 
%contrast to numerical predictions which showed unidirectional liquid motion. 

{
Most theoretical and numerical electroosmosis works with nonuniformly charged walls employ Neumann boundary conditions, that is, the normal component of the electric field to a wall is fixed by the commensurate charge distribution, see for instance \citet{ajdari1995,Kamsma2023prl,Kamsma2023}. 
On the other hand, most experimental works and their accompanying theories, employed Dirichlet boundary conditions, that is they fix
the electric potential on the walls, e.g.  \citep{Cahill2004,Gonzalez2008}. In the literature there is no established argument to justify the use of one or the other. They lead to different predictions and it is not apparent if there is any relationship between
them. A fundamental physical justification for the presence of the unidirectionally observed flow (termed the zero mode here) is still lacking. There is no scaling argument in the literature that would classify the observed flows in a universal manner. The approximations
involved usually require small frequencies, $\omega \ll D k\kappa$, where $\kappa$ is the inverse Debye length and 
$k$ is the excitation wavenumber to be defined below, cf. \citep{Ramos2005,Gonzalez2008}. 
It is common practice in the literature to separate
the spatial domain into several adjacent regions over which different equations apply. This also requires the introduction of different material
parameters in each region not necessarily known \emph{a-priori}.
The current state-of-the art is based on different configurations, boundary conditions, liquids etc., making it
hard, if not impossible to reach a unified view of the observed effects.  In addition, 
velocity profiles that appeared in the published literature \citep{ehrlich1982}, are 
not associated with a vector quantity which, upon reversal, would also reverse the direction of the 
(zero mode) velocity, creating an ambiguity on the nature of 
the mechanism that drives the flow. 
We will address all these issues in turn, and we will unveil the universal aspects of the flow, as 
described below.

In this paper we develop a hydrodynamic theory for the unidirectional motion of an electrolyte when the bounding walls
(of a semi-infinite space, channel or cylindrical capillary) are insulating and carry a traveling wave charge distribution $\sigma = \sigma_0 e^{i(\mathbf{k} \cdot \mathbf{x} - \omega t)}$  (Neumann boundary conditions for the Poisson equation), where $\mathbf{k}$ and $\omega$ are the 
wavevector and angular frequency, respectively, of the plane wave. This is in contrast to the standard route taken in the literature where (some type of) Dirichlet boundary conditions are adopted \citep{Cahill2004,Ramos2005}. After averaging over the period of oscillations, we show that the non-vanishing electrolyte velocity 
profiles (that we term the zero-mode) are self-similar and only depend on three groups of dimensionless parameters. With this 
realization, it is possible to draw conclusions for the system behavior based on different liquids, charge distributions, 
electric field amplitudes, frequencies etc., by just performing a single experiment. 
%The self-similar profiles can be well described by a simple expression, or its modification, originally derived by \citet{ehrlich1982}. 
We repeat the above development for the case of Dirichlet boundary conditions for the Poisson equation 
(traveling wave electric potential $\phi = \phi_0 e^{i(\mathbf{k} \cdot \mathbf{x} - \omega t)}$ at the wall)
and show that self-similar profiles also exist but for different dimensionless groups. In particular, while the Neumann problem requires that the excitation frequency $\omega$ is normalized by
the Debye time scale $\tau_D$, where, $\tau_D^{-1} = D \kappa^2$, the Dirichlet problem frequency
%depends on the excitation wavenumber $k = |\mathbf{k}|$ \emph{and} on system configuration: the frequency in the semi-infinite space is normalized
is normalized by 
$\tau_{k\kappa}^{-1} = Dk\kappa$ (semi-infinite space) and by $\tau_k^{-1} = Dk^2$ (channel case).  Thus, Neumann boundary conditions seem to furnish a more robust and consistent behavior, relative to its Dirichlet counterpart, at all configurations since its dimensionless group does not depend on geometry. We note that this predominance of Neumann versus their Dirichlet counterparts has already been pointed-out in the literature of electrokinetic energy conversion and its efficiency in 
nanofluidic channels by matching theory \citep{vanderHeyden2006} with experiment \citep{vanderHeyden2007}. 
In addition, variational formulations employ the Neumann conditions as \emph{natural} boundary conditions in general and in the solution of the Poisson-Boltzmann equation in particular \citep{Clarke2015}. 
We perform finite-element numerical
simulations of the full equations (with the Neumann boundary conditions) and find good agreement with their exact counterparts in an order-of-magnitude basis and in their
general trend.
A well-known principle, called Stokes' Rule \citep[p. 258]{Zauderer1989}, then explains the relationship between
the Neumann and the Dirichlet results (cf. Appendix \ref{sec: kscaling}). 
%and that
%they collapse to the curves of a simple theoretical expression derived by \citet{Ramos2005}. 

We introduce simple theoretical expressions (theory fits) that reproduce the self-similar profiles, and only 
depend on a single fit parameter $\beta$,
which could thus provide a rapid test of consistency between our
theory and future experiment. 
We show that our
order-of-magnitude estimates of the liquid velocity derived with the Neumann theory might agree well with experiment. This is to be contrasted with theories, directly or indirectly based on Dirichlet conditions, which provide 
much higher estimates in comparison to the measurements of their accompanying 
experiments, cf. \citet{Cahill2004,Ramos2005}. 

We show that the fluid velocity with Neumann conditions becomes 
prominent by decreasing the transverse size of the device, relative to the direction of the velocity, or by increasing the excitation wavelength,
and thus could be used for unidirectional electrolyte transport in thin and long devices. This behavior was numerically established before, cf. \citet[Fig. 6(f)]{Liu2018}. 
%The fluid velocity
%with Dirichlet boundary conditions, however, displays \emph{exactly the opposite behavior}. 

The traveling-wave wall charge-induced unidirectional velocity 
is a \emph{zero} or massless or soft or 
Goldstone mode \citep{Goldstone1961}.
The existence of a zero mode is associated with broken continuous symmetries \citep{Negele1988} and its presence here is not surprising as it is well known that it 
arises in problems involving quadratic nonlinearities, such as the Kuramoto-Sivashinskii equation cf. \citep{malomed1992,Kirkinis2014SH}. Here the quadratic nonlinearity is due to the electric body force
in the momentum equation. 
%----------------------------------------------------------------------------------------
% there is a problem with bibtex calling Kirkinis2014b, so I retyped and renamed it 2014SH
%-----------------------------------------------------------------------------------------

Certain results from the literature are recovered as special cases of our formulation (for instance, we recover the \citet{Ramos2005}
theoretical expression for the unidirectional velocity) and we resolve certain paradoxes that 
appeared before (for instance, that the traveling wave electroosmosis solution of \citet{ehrlich1982} is singular).
We derive \emph{general} formulas, for the `slip' or average velocities that only depend on the form taken by the electric potential and the geometry of the system. This is useful since these formulas can be adopted `as is' even when different boundary conditions of the electric problem are employed.

In this paper we invoke the Debye-Falkenhagen approximation \citep{Bazant2004}, a necessary step towards retaining the time-dependence of the Nernst-Planck equation. We are unaware of any study of its range of validity; we thus provide 
such a detailed analysis, which can only be carried-out on an \emph{a-posteriori} basis. 
We show that the approximation is superior to its 
Debye-H\"uckel counterpart and its validity increases with increasing excitation frequency $\omega$.
To avoid any confusion, we emphasize that the term ``nonlinear'', employed in this paper, refers to the form of the governing partial differential equations endowed with a quadratic nonlinearity in the momentum equation and 
\emph{is not} associated with retaining the Boltzmann form of the species concentrations. 
}

This paper is organized as follows. 
Section \ref{sec: traveling} introduces the governing equations and boundary conditions of a \emph{nonlinear} formulation of the electroosmosis problem when the electrolyte is driven
by traveling wave charge distributions on the channel walls (or driven by traveling wave wall electric potentials). 
In section \ref{sec: space}, we consider the electrolyte filling the semi-infinite space $z>0$, driven by a traveling wave 
charge distribution at the wall $z=0$ with wavenumber $k$. We show that the \emph{zero mode} velocity at $z=\infty$ 
(customarilly refered to in the literature as the ``slip velocity'') is self-similar with respect to 
three dimensionless groups and that a simple expression can be employed 
to describe these profiles. 
In section \ref{sec: infscaling} we provide estimates of the velocity magnitude for standard experimental 
parameters,  and show that they agree well, in order-of-magnitude, with existing
measurements, in contrast to their Dirichlet theoretical counterparts. 

In section \ref{sec: channel} 
we consider traveling charge distributions on both walls of a channel giving rise to a \emph{zero mode} (that is, 
unidirectional and time-independent)
velocity field. This configuration choice is dictated by results showing that 
the ideal and optimal geometry
requires in-phase and symmetrical electrode arrays with respect to the channel center
\citep{Yeh2011}. 
The zero mode velocity averaged over the channel width is again self-similar and is compared to its 
(also self-similar) Dirichlet counterpart.  The flow velocity increases by reducing the channel size, reaching 
a steady value. In section \ref{sec: cylinder} we reconsider the nonlinear electroosmosis problem but now in a cylindrical capillary 
whose wall carry traveling wave charges. We reach similar conclusions to the channel case regarding
the presence of the zero mode and its various limits. 
%As the optimal zero mode velocity is expected to arise when the 
%wall excitations are identical and of the same phase (see \citep{Yeh2011} for the case of traveling wave voltages), 
%we have adapted our formulation to these conditions. 

In section \ref{sec: numerical} we employ numerical simulations of the Poisson-Nernst-Planck-Navier-Stokes
system (in the low P\'eclet number approximation) to obtain the zero mode velocity and compare it with its
exact counterparts developed in the previous sections. Details of the numerical scheme employed are 
delegated to a Supplementary Materials addendum.

We conclude this paper with a number of Appendices. These include, an analysis of the validity of the
Debye-Falkenhagen approximation, the solution of the Poisson-Nernst-Planck problem for Dirichlet boundary 
conditions and a discussion of the $k$-dependence of the various velocities derived in the main body of
this article.
We show that if the zero-mode velocity is taken into account in the Nernst-Planck equation, the observables are not significantly affected for moderate values of the P\'eclet number
cf. Appendix \ref{sec: Pe}. In Appendix \ref{sec: E&M} 
we show that the traveling wave electroosmosis solution of \citet{ehrlich1982} is singular and that the commensurate zero mode velocity field is non-unique.

\begin{figure}
\vspace{5pt}
\begin{center}
\includegraphics[height=1.7in,width=5in]{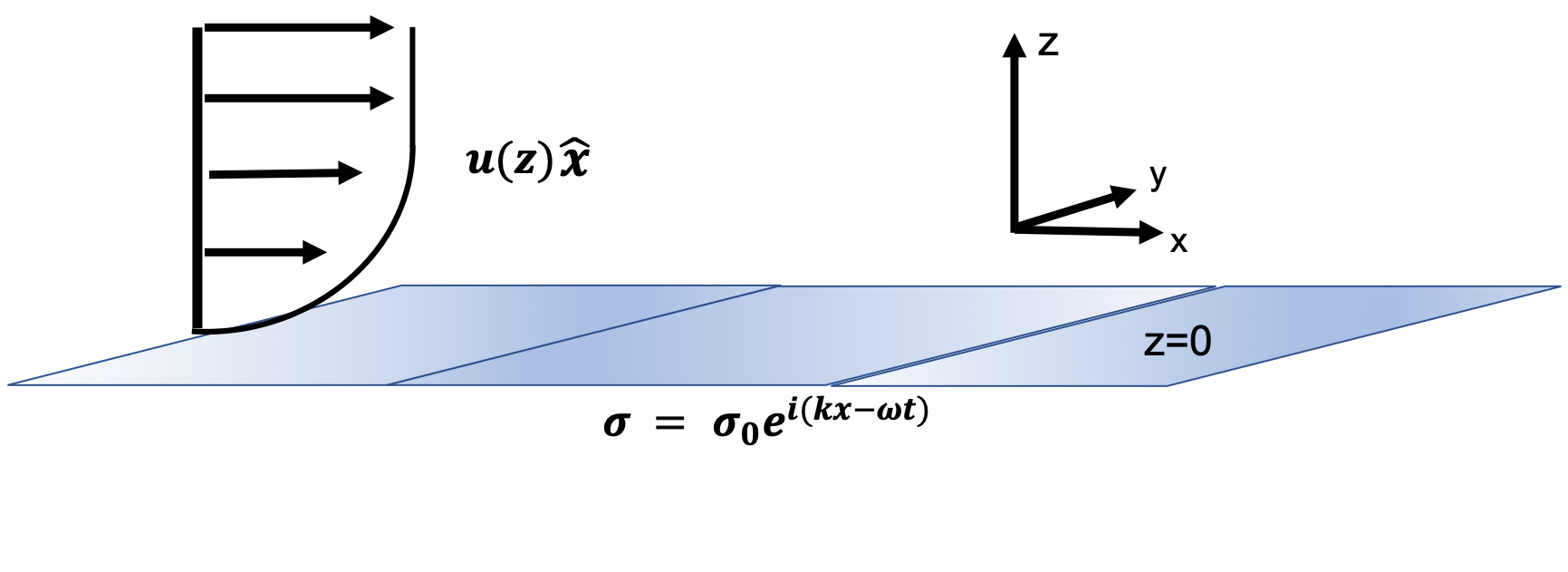}
\vspace{-0pt}
\end{center}
\caption{Wall traveling wave charges give rise to a nonlinear body force and torque (see \rr{psinl}) in a $1:1$ electrolyte lying in the semi-infinite space $z>0$, leading to the appearance of 
a unidirectional fluid velocity in the $\hat{\mathbf{x}}$ direction, parallel to the wall, that is quadratic 
with respect to the associated electric field and that does not vanish after averaging over the charge period of oscillation. 
\label{space1wave}  }
\vspace{-0pt}
\end{figure}

\section{\label{sec: traveling}Governing equations of traveling wave electroosmosis}

Consider the application of traveling-wave charges 
\be \label{sigmatravel}
\sigma(x,t) = \sigma_0 e^{i (kx -\omega t)}
\ee
on an insulating wall, temporarily identified with the plane $z=0$, cf. figure \ref{space1wave},
with real frequency $\omega$ and real wave-vector $\mathbf{k} = k \hat{\mathbf{x}}$.  
We consider a $1:1$ electrolyte where each concentration species is $c_\pm =c_\infty + \delta c_\pm$ containing 
a perturbation $\delta c_\pm$ superposed on its uniform counterpart $c_\infty$. Thus, the bulk charge distribution is $\rho = e(\delta c_+ - \delta c_-)$
while the salt distribution is $ s\sim 2 ec_\infty$, to leading order, where $e$ is the proton charge. 
{
This implies that $s$ is also accompanied by a perturbation $e(\delta c_+ + \delta c_-)$, which however is never invoked in this paper but it
can be computed, need be.}
 The validity of this approximation can only 
be examined \emph{a-posteriori} and in conjunction to the reductive form of the Nernst-Planck equation, as detailed below. We carry-out such an analysis in Appendix \ref{sec: validity}. 
%There are three processes by which affect the charge distribution can change within a fluid element. First, when the liquid moves as a whole
%(with velocity $\mathbf{v}$) and its composition remains unchanged. Second, due to the transfer of the charge (on the molecular level)
%from one part of the liquid to another (the irreversible process called diffusion), and third, charge transfer due to the presence of an electric field. 

Invoking the aforementioned approximation and the Poisson equation for the electric potential $\phi$
\be \label{phit}
\nabla^2 \phi = - \frac{\rho}{\epsilon},  
\ee
where $\epsilon$ is the dielectric constant, 
the evolution of charge distribution %(due to all three aforementioned processes),
\be  \label{rhotgeneral}
\partial_t \rho = D\left[ \nabla^2 \rho + \frac{e}{k_BT} \nabla \cdot \left( s \nabla \phi \right) \right] - \mathbf{v} \cdot \nabla \rho,
\ee
 reduces to the Debye-Falkenhagen equation \citep{Bazant2004}
\be \label{rhot}
\partial_t \rho = D\left[ \nabla^2 \rho -\kappa^2 \rho \right] -\partial_x \rho \partial_z \psi + \partial_z\rho \partial_x \psi. 
\ee
Here, $\mathbf{v}$ is the electrolyte velocity, $D$ is the charge diffusion coefficient (assumed to have the same value for both species), $k_BT$ the thermal energy
and $\psi = \psi(x,z, t)$ is the streamfunction for a two-dimensional incompressible liquid, whereby
\be \label{psi0}
\mathbf{v} = (u, 0, w) = \left(\frac{\partial \psi}{\partial z}, 0, -\frac{\partial \psi}{\partial x}\right),
\ee
and 
\be \label{kappa}
\kappa = \left[ \frac{2e^2 c_\infty}{\epsilon k_BT} \right]^{\frac{1}{2}}
\ee
is the inverse Debye length. 
%{Eq. \rr{kappa} is written in terms of the compact notation $c_\infty$ for the 
%charge distribution (following the definition of bulk charge in \citep{ajdari1995}). To obtain the more customary expression, replace $c_\infty$ by $e n^\infty$
%where $n^\infty$ is the concentration of charge species. }
%Thus, the above approximation is not limited to small Debye lengths. 
How this approximation relates to its classical Debye-H\"{u}ckel counterpart (where $\rho = -\epsilon \kappa^2 \phi$ everywhere), 
is discussed in Appendix \ref{sec: approximation}. 

Even in the absence of the 
last two (nonlinear advective) terms, Eq. \rr{rhot} leads to a \emph{non-equilibrium} charge distribution $\rho$, where the 
charge changes both in time and in space, in contrast to the essentially equilibrium route taken in the literature, cf. \citep{Probstein1994}. 
Eq. \rr{rhot} is not new and has  been derived earlier, for instance, in \citet{Cahill2004,Mortensen2005} following the older results of 
\citet{ehrlich1982}. 

{ 
In the main body of this paper we will adopt the low P\'eclet number limit, thus reducing \rr{rhot} to 
\be \label{rhot2}
\partial_t \rho = D\left[ \nabla^2 \rho -\kappa^2 \rho \right] .
\ee
We justify the small P\'eclet number approximation in Appendix \ref{sec: linearization}}. We show in Appendix
\ref{sec: Pe} that retaining the advection terms in \rr{rhot}, based on the zero velocity mode only, does
not lead to a significant change in the observables for moderate values of the P\'eclet number.

Finally, the streamfunction $\psi$ satisfies 
\be \label{psinl}
\rho_l\partial_t \nabla^2 \psi = \eta \nabla^4 \psi  + \partial_x \rho \partial_z \phi - \partial_z\rho \partial_x \phi,
\ee
where $\rho_l$ is the liquid electrolyte density and $\psi(x, 0, t) = 0 = \partial_z \psi(x,0, t)$ (the no-slip boundary condition) assuming temporarilly that
the solid boundary is identified with the plane $z=0$. 
If the flow extends to infinity then its velocity is considered to have a finite value there. 

All results of the present section depend on the (weakly) nonlinear character of the nonlinear torque (last two terms) in 
Eq. \rr{psinl}. Although, in general, the commensurate velocity field vanishes when averaged over the period of oscillation
of the applied electric field, there are circumstances where this is not so. This is due to the presence of a zero mode that 
arises due to constructive interference of the charge and potential excitations arising in \rr{psinl}. The nonlinear
effect described in \rr{psinl} vanishes if the charge and the potential do not vary in the direction
parallel to the wall (the $x$-direction). {Likewise, in the $\omega \equiv 0$ case, the nonlinear torque in \rr{psinl} 
vanishes, on account of the connection $\rho= -\epsilon \kappa^2 \phi$ (the Debye-H\"uckel approximation), or equivalently, the electric body force is the gradient
of a scalar field and thus only affects the pressure distribution. }

\subsection{\label{sec: bcs}Boundary conditions}
{In this paper we will predominantly consider Neumann boundary conditions for the Poisson equation \rr{phit}, 
at a solid wall carrying a surface charge $\sigma$ as in \rr{sigmatravel},
where
\be \label{N1}
 \hat{\mathbf{n}} \cdot \nabla  \phi  = -\frac{\sigma}{\epsilon}, \quad \textrm{at a charged wall}
\ee
with the unit normal vector $\hat{\mathbf{n}} $ pointing into the liquid. We will consider also the case
where the current of bulk charge $\rho$ vanishes at wall, that is,
\be \label{jn}
 \mathbf{J}\cdot \hat{\mathbf{n}} \equiv  - D\left[ \nabla \rho + \frac{e}{k_BT} s \nabla \phi  \right] \cdot \hat{\mathbf{n}}=0
 , \quad \textrm{at a charged wall}.
\ee
It is more convenient to replace \rr{jn} with a simpler expression. The same approximation that allowed the reduction of \rr{rhotgeneral} into \rr{rhot}, leads the vanishing current condition \rr{jn}
at the wall, to become a boundary condition for the bulk charge $\rho$
\be \label{N2}
 \hat{\mathbf{n}} \cdot \nabla  \rho = \kappa^2 \sigma, \quad \textrm{at a charged wall}. 
\ee
Finally, appropriate conditions must be set at infinity if the domain is unbounded. 
 
We should stress here that the system of equations (the Poisson-Nernst-Planck system) \rr{phit} and \rr{rhot}
with the Neumann boundary conditions does not have a solution when $k\equiv 0$. 
%Physically, this value corresponds to injection of an infinite amount of energy into the system. 
Mathematically, 
the 
electric potential cannot satisfy all boundary conditions. However, the devices we have in mind are, say, at most
$10$ cm long, with a wavenumber $k=10 \textrm{ m}^{-1}$. Past experiments on traveling wave electroosmosis
employed electrode arrays with $k\sim 10^4 \textrm{ m}^{-1}$ \citep{Ramos2005} or even larger \citep{Cahill2004}. We will thus operate within the interval of physically realistic wavenumbers
$k\in (10^1, 10^5) \textrm{ m}^{-1}$,
as we stated in table \ref{tab: table1}.
%
%On the other hand, the same system of equations but with Dirichlet boundary conditions (fixing the electric potential at a wall), has a well-behaved solution. We solve this type of system in the Appendix. However, 
%we should recall that a recurrent theme in traveling wave electroosmosis is the overestimation of the experimental results by the theory (based on essentially Dirichlet boundary conditions). As
%will become
%soon apparent, the Neumann boundary conditions are more physically realistic and provide estimates of the electrolyte velocity that lie closer to experiment than their Dirichlet counterparts. 
%

The notation used  in \rr{psinl} implies that $\rho$ and $\phi$ are real fields. In the subsequent
discussion we will employ the same notation to denote complex fields from now on.

\subsection{\label{sec: scales}Velocity scales}
We will employ the following velocity scales
\be \label{u01}
u_0 = \frac{1}{2}\frac{\epsilon E^2}{\eta \kappa}, \quad
u_1 = \frac{\epsilon \kappa \phi_0^2}{2\eta}, 
\ee
where $$E \equiv \frac{\sigma_0}{\epsilon}$$ is the electric field amplitude due to the
interfacial charge distribution $\sigma_0$ (Neumann problem) 
and $\phi_0$ a characteristic scale for the electric potential (Dirichlet problem). 
We must reiterate here that the electric field $E$ is induced by the charged walls in the \emph{boundary-guided} traveling-wave electroosmosis effect studied in the present paper, and differs substantially to the electric field (potential difference) applied in a capillary in standard electroosmosis, e.g. \citet{ajdari1995}. 
Dimensionless groups
will be stated in raw form to avoid introducing new notations.

For brevity we will use the phrases ``Neumann velocities'' or ``Dirichlet velocities'' to denote 
velocities obtained by solving the corresponding Poisson equation with Neumann or Dirichlet
boundary conditions, respectively. 
}

\section{\label{sec: space}Traveling wave electroosmosis in a semi-infinite space}

We consider the configuration displayed in Fig. \ref{space1wave}. Traveling wave surface charges $ 
\sigma(x,t) = \sigma_0 e^{i (kx -\omega t)}$
are applied on the wall $z=0$
bounding an $1:1$ electrolyte lying in the semi-infinite space $z>0$.
The boundary conditions satisfied by the bulk charge $\rho$ and electric potential $\phi$ at a solid surface lying at $z=0$
with the unit vector $\hat{\mathbf{n}}=\hat{\mathbf{z}}$ are, from \rr{N1} and \rr{N2}
\be \label{phirho}
\partial_z \phi (x,z = 0, t) = -  \frac{\sigma(x, t)}{\epsilon}, \quad \partial_z \rho (x,z = 0, t) = \kappa^2 \sigma(x, t). 
\ee

\begin{table}
  \begin{center}
\def~{\hphantom{0}}
 % \begin{tabular}{|c|c|c|c|}
\begin{tabular}{lcl}
\textrm{Quantity}&
\textrm{Value}&
%\multicolumn{1}{c}{\textrm{Decimal}}&
\textrm{Definition}\\
$\omega\; (\textrm{rad} \cdot \textrm{sec}^{-1})$ & arbitrary & wall traveling wave charge frequency\\
$\Delta = \sqrt{\frac{2 D}{\omega}} \; (\textrm{cm})$ & $$ & charge penetration depth (cf. \rr{kdelta2}) \\
$k \; (\textrm{m}^{-1})$ & $10^1-10^5$ & wall charge wavenumber\\
$\kappa \; (\textrm{m}^{-1})$ & $10^2-10^8$ & inverse Debye length \\
$K=\sqrt{k^2 + \kappa^2} \; (\textrm{m}^{-1})$ & $$ &\\
$p=\frac{1+i}{\Delta} \; (\textrm{m}^{-1})$ & $$ & charge oscillating complex wave number (cf. \rr{kdelta2}) \\
$P = \sqrt{p^2 - k^2-\kappa^2} \; (\textrm{m}^{-1})$ & $$ & charge complex wave number (cf. \rr{kdelta2}) \\
$\eta$ ($\textrm{kg}\:\textrm{m}^{-1}\textrm{sec}^{-1} $)    & $10^{-3}$         & electrolyte dynamic viscosity \\
$\nu$ ($\textrm{m}^{2}\textrm{sec}^{-1} $)    &     $10^{-5}$    & electrolyte kinematic viscosity \\
$2h$ (m) & $10^{-2}-10^{-5}$ & channel width\\
$a$ (m) & $10^{-2}-10^{-5}$ & capillary radius\\
$u$ (m/sec) & $10^{-4}$ & horizontal (zero mode) velocity component\\
$u_0,u_1$ (m/sec) &  & horizontal velocity scales, cf. Eq. \rr{u01}\\
$D\; (\textrm{m}^{2}\textrm{sec}^{-1} )$  & $10^{-9}$ & Diffusion coefficient for electrolyte charges\\
$\sigma, \sigma_0 \; (\textrm{C/m}^2) $  & $$ & wall charge distribution\\
$E$ (V/m)& $10^5$ & Electric field\\
$\epsilon$ (F/m) &$ 7\times 10^{-10}$& Dielectric constant\\
$\phi , \phi_0$ (V) & $1-6$  & electric potential: $\mathbf{E} = -\nabla \phi$ \\
$\psi $  & $$ & streamfunction, cf. \rr{psi0}\\
$\rho, s$  & $$ & bulk charge and salt distribution\\
$\mathcal{P}$  & & hydrodynamic pressure\\
$\rho_l$  & & ionic liquid density
\end{tabular}
\caption{\label{tab: table1}%
Definitions of wavenumbers and parameter values. $\rho, k, K, \sigma$ and $\kappa$ as in  \citet{ajdari1995}.  Our $k$ and $K$ correspond to the $q$ and $Q$ of \citet{ajdari1995}.
$p$ and $\Delta$ correspond to the $k$ and $\delta$ of \citet[\S 24]{Landau1987}.
}
  \end{center}
\end{table}

Assuming $\rho = \rho(z) e^{i (kx -\omega t)}$, Eq. \rr{rhot2} reduces to $ \rho_{zz} + \left[i\omega/D -\kappa^2 -k^2\right] \rho =0$.
To avoid clutter, we have introduced the notation
$\rho_z\equiv \partial_z \rho$ etc. 
Thus, the charge distribution reads
\be \label{rhoxzt1}
\rho(x,z,t) = -\frac{i\sigma_0\kappa^2}{P} e^{i(Pz +kx -\omega t)}, 
\ee
where we assumed a vanishing bulk charge at infinity and introduced the notation
\be \label{kdelta2}
P \equiv P_1+iP_2 = \sqrt{p^2 -k^2 -\kappa^2}, \quad
i\omega = D p^2, \quad  p = \frac{1+i}{\Delta}, \quad \Delta = \sqrt{\frac{2D}{\omega}}.  
\ee

The notation employed in Eq. \rr{kdelta2} for penetration depth $\Delta$ and complex wavenumber $p$ is analogous 
to \citep[p.84]{Landau1987}. These quantities here however refer to the wavelike
form of the charge distribution away from a charged boundary, rather than the wavelike form of the velocity field away
from a no-slip wall.  
%Thus, in the absence of charge modulation ($k=0$) and double layer ($\kappa=0$) the charge diffuses away from the wall
%with penetration depth $\Delta$ (the inverse of the imaginary part of $p$) and oscillates with period $2\pi/\Delta$ away from the wall (the real part of $p$). This picture is also 
%present when $\kappa$ and $k$ are non-zero. 
The penetration depth is now determined by the 
inverse of the imaginary part of $P$ (taken here to be positive) and there is an oscillation whose wavevector is the real 
part of $P$. In other words, if $P_1$ and $P_2$ denote the real and imaginary parts of $P$ in \rr{kdelta2} ($P = P_1+iP_2$), where $P_i$ are real and $P_2>0$ then
\be \label{P12}
P_{1,2} = \frac{1}{\sqrt{2}} \sqrt{\sqrt{\left(\kappa^{2}+k^2\right)^{2}+\frac{\omega^{2}}{\mathit{D}^{2}}}\mp (\kappa^{2}+k^2)},
\ee
where the minus/plus sign corresponds to the real/imaginary part of $P$. Thus, the penetration depth is not determined by $2\pi/\kappa$
but by the length scale $2\pi/P_2>0$. 
With this notation, the charge distribution has the form
\be \label{rhoxzt}
\rho(x,z,t) =- \frac{i\sigma_0\kappa^2}{P_1+iP_2} e^{-P_2z}e^{i(P_1 z +kx -\omega t)}. 
\ee
Thus, in addition to the exponential decay away from the wall, there is a plane wave whose 
wavevector is not parallel to the surface charge distribution wave-vector (it is not parallel to the $x$-axis) 
in Eq. \rr{sigmatravel}
but lies in the direction $k \hat{\mathbf{x}} + P_1 \hat{\mathbf{z}}$.

\begin{figure}
\vspace{5pt}
\begin{center}
\includegraphics[height=2.6in,width=4in]{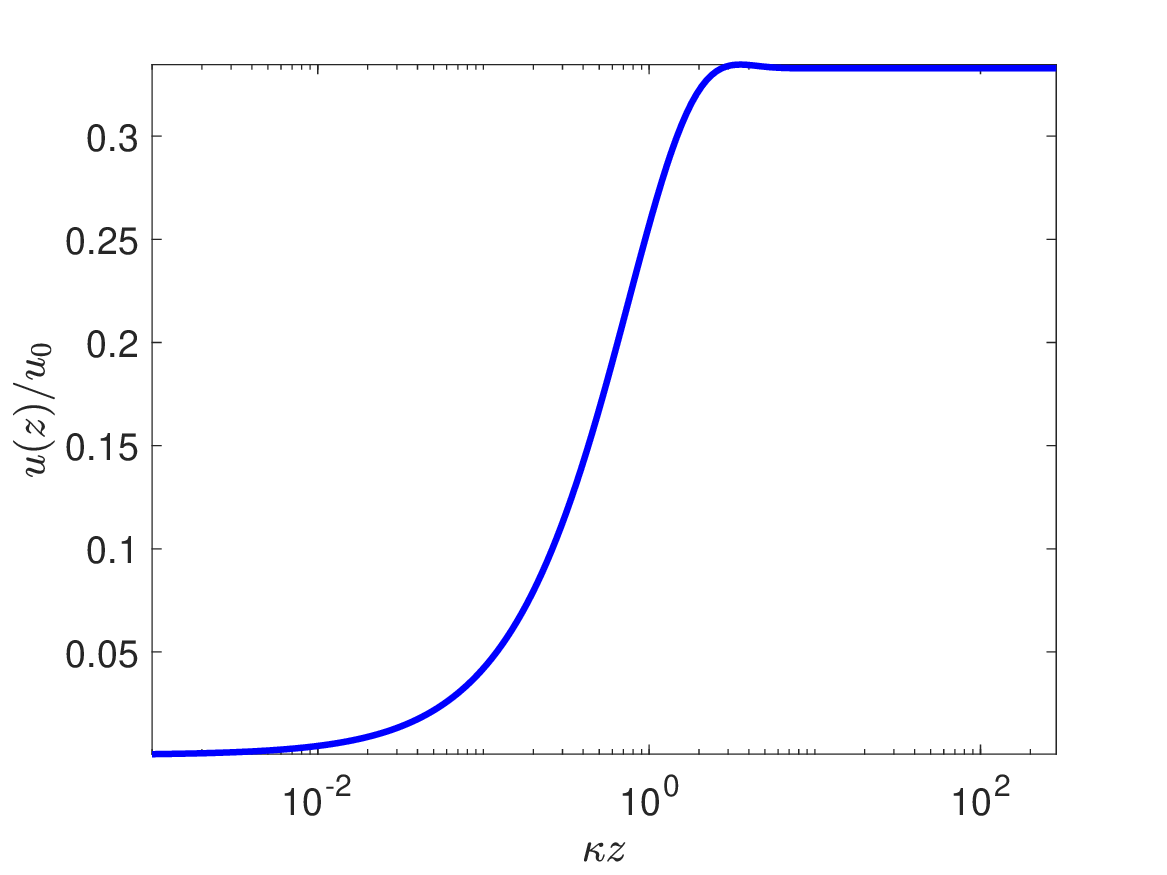}
\vspace{-0pt}
\end{center}
\caption{Boundary layer of thickness (= penetration depth, see Eq. \rr{kdelta2}) of $O(P_2^{-1})\sim \kappa^{-1}$ for large $\kappa$ employing 
\rr{uzinf}. 
Beyond the boundary layer the liquid velocity rapidly acquires a constant value. 
$\kappa^{-1}, D, \eta$ are the Debye length, charge diffusion coefficient
and liquid viscosity, respectively. Here we have taken $k =10^5 \textrm{ m}^{-1}$, $\kappa =10^7 \textrm{ m}^{-1}$
$D = 10^{-9}\textrm{ m}^{2}/$sec and $\omega = D\kappa^2$. $u_0$ was defined in \rr{u01}.
\label{uz}  }
\vspace{-0pt}
\end{figure}

Similarly, assuming $\phi = \phi(z) e^{i (kx -\omega t)}$, Eq. \rr{phit} reduces to $ \phi_{zz} -k^2 \phi =\frac{i\sigma_0\kappa^2}{\epsilon P} e^{iPz}, $ subject  to the 
boundary condition \rr{phirho} with surface charge \rr{sigmatravel} 
and $\phi =0$ at infinity (the alternative boundary condition of zero electric field at infinity leads to an identical result). 
Thus, the potential distribution reads
\be \label{phixzt}
\phi(x,z,t) = \frac{\sigma_0}{\epsilon} \left( 1 + \frac{\kappa^2}{P^2 +k^2} \right)  \frac{e^{-|k|z}}{|k|} e^{i(kx -\omega t)}
+\frac{1}{\epsilon (P^2 +k^2)} \rho, 
\ee
with $\rho$ given by \rr{rhoxzt}. Note that, as we mentioned in the previous section, setting $k=0$, the boundary value problem does not have a solution since the electric potential $\phi$ cannot
satisfy both boundary conditions. This is then reflected in the divergence of expression \rr{phixzt} as $k\rightarrow 0$. For the purposes of this paper, however, where the physically realistic wavenumber $k$ lies in the interval $(10^1, 10^5) \textrm{ m}^{-1}$, 
this limit is never attained. 

Also, see Appendix \ref{sec: approximation} for a comparison between the result \rr{phixzt} and
its Debye-H\"{u}ckel counterpart.

It is now clear that the nonlinear torque (last  two terms in \rr{psinl}) involves the harmonics
$e^{\pm i \theta}$ and $e^{\pm 2i\theta}$ where $\theta = kx-\omega t$ and the commensurate 
streamfunction $\psi$ satisfying \rr{psinl} will also be composed of the same  harmonics. 
Thus, the velocity vanishes when averaged over the period $2\pi/\omega$ of oscillations. 

There is however a part of the nonlinear torque in \rr{psinl} that is time-independent and \emph{does not vanish} 
when averaged over the period of oscillations. This is caused by constructive 
interference of the various harmonics expressing  the charge and potential distribution in \rr{rhoxzt} and \rr{phixzt}, respectively
and gives rise to \emph{unidirectional} pumping of liquid, its direction determined by the propagation direction of the 
charge distribution lying on the walls. We now investigate this zero mode. 

%A useful form is the one obtained by also averaging over $z$ from 
%the solid surface (located at $z=0$) to infinity. It reads 
%\be
%\langle \langle N \rangle \rangle = -\frac{\kappa^{2} \sigma_0^{2} P_{1} \left(\kappa^{2}+k^2+2 q P_{2}+P_{1}^{2}+P_{2}^{2}\right)}{\epsilon  \left(P_{1}^{2}+P_{2}^{2}\right) \left(k^2+2 q P_{2}+P_{1}^{2}+P_{2}^{2}\right)}
%\ee
%and the double angle brackets denote averaging over a period of oscillation $2\pi/\omega$ and over space.

Let $\psi = \psi(z)$, and thus the velocity $ \mathbf{v} = u(z) \hat{\mathbf{x}}$ as in \rr{psi0}, and consider both $\psi$ and $u$ to be real fields, cf. Fig. \ref{space1wave}. In Appendix \ref{sec: nltorque} we show that after averaging over the period of oscillation, the vorticity equation \rr{psinl} reduces to 
\be \label{upz}
\partial_zu(z) = \frac{i\epsilon k}{4\eta}  \left( \phi^*\phi_z - \phi \phi_z^*\right),
\ee
where a star on $\phi$ denotes the complex conjugate of \rr{phixzt}. 
%This is also the vorticity. 
It is clear that the complex exponentials $e^{i(kx -\omega t)}$ have canceled-out. This is the central formula of this paper.
It states that the shear stress in a viscous electrolyte is given by the expression on the right-hand side (multiplied by viscosity) which resembles a probability current for a Schr\"odinger equation \citep{Morse1953}.
%in the parabolic approximation\citep{Zauderer1989}. 
We also verify the validity of \rr{upz} by starting directly from the time-dependent Stokes equations see Appendix \ref{sec: ns}.
 
It is thus easy to determine the horizontal velocity component $u(z)$. 
Employing the complex form of the field $\phi$ in \rr{phixzt} we solve \rr{upz} subject to the boundary conditions
\be \label{upzbc}
u(z=0)=0, \quad u(z=\infty) = \textrm{finite}. 
\ee
Thus, the zero mode velocity $u(z)$ from \rr{upz} with \rr{phixzt} becomes 
\be \label{uzinf}
u(z) = \frac{i\epsilon k}{4\eta} \int_0^z \left( \phi^*\phi_z - \phi \phi_z^*\right) dz.
\ee
The zero mode \rr{uzinf} inherits the boundary-layer structure of the field $\phi$ as displayed in Fig.  \ref{uz}. A boundary
layer of thickness $O(P_2^{-1})\sim \kappa^{-1}$ for large $\kappa$ separates the no-slip region at the wall to the 
constant value acquired by the velocity away from the wall.

In the semi-infinite space formulation of the present section, the observable of interest is the value taken by the electrolyte velocity zero mode \rr{uzinf} at $z=\infty$. 
It reads
\be \label{uzpolarabs}
\frac{u(\infty)}{u_0} =
\pm \frac{\kappa^{3} \left(2 R^{3} \cos \left(3 \Theta \right)-4 \sin \! \left(2 \Theta \right) R^{2} |k| +2 R (\kappa^{2} - k^2) \cos \left(\Theta \right)+\kappa^{2} |k| \cot \! \left(\Theta \right)\right)}{2 R^{2} \left(2 R^{2} k^{2} \cos \! \left(2 \Theta \right)-4R|k| (R^2 + k^2) \sin \! \left(\Theta \right) -R^{4}-4 R^{2} k^{2}-k^{4}\right)},
\ee
where the plus sign gives the velocity for positive $k$ and the negative sign for negative $k$
%\be \label{uzpolar}
%u(\infty) = \frac{\kappa^{3} \left(R^{3}\sin \! 4 \Theta-R \left(R^{2}-\kappa^{2}+k^2\right) \sin \! 2 \Theta  -q \left(2 R^{2}-\kappa^{2}\right) \cos \! \Theta +2R^{2} q \cos \! 3 \Theta  \right)}{2 R^{2} (R^{2} k^2\sin \! 3 \Theta -(R^{4}+5 R^{2} k^2+k^4) \sin \! \Theta   -2 R q (R^{2}+k^2) (1-\cos \! 2 \Theta) )}
%\ee
and, following 
\citep{ehrlich1982}, the horizontal velocity $u$ was scaled by $u_0$, cf. \rr{u01}. 
%\be \label{u0}
%u_0 = \frac{1}{2}\frac{\epsilon E^2}{\eta \kappa},
%\ee
%where $E = \frac{\sigma_0}{\epsilon}$ is the electric field amplitude due to the
%interfacial charge distribution $\sigma_0$. 
Here 
$R$ and $\Theta$ are the amplitude and phase of the complex wavenumber $P = R e^{i\Theta}$ defined  in \rr{kdelta2} and given
explicitly by 
\be \label{RTheta}
R = {\left[\left(\kappa^{2}+k^2\right)^{2}+\frac{\omega^{2}}{\mathit{D}^{2}}\right]}^{\frac{1}{4}}, \quad 
\Theta = -\frac{1}{2}\arctan \! \left[\frac{\omega}{\mathit{D} \left(\kappa^{2}+k^2\right)}\right]+\frac{\pi}{2}. 
\ee
In Appendix \ref{sec: E&M} we discuss the relation of the velocity \rr{uzpolarabs} with its counterpart
that appeared in the past literature \citep{ehrlich1982}.

\begin{figure}
\vspace{5pt}
\begin{center}
\includegraphics[height=2.8in,width=5.6in]{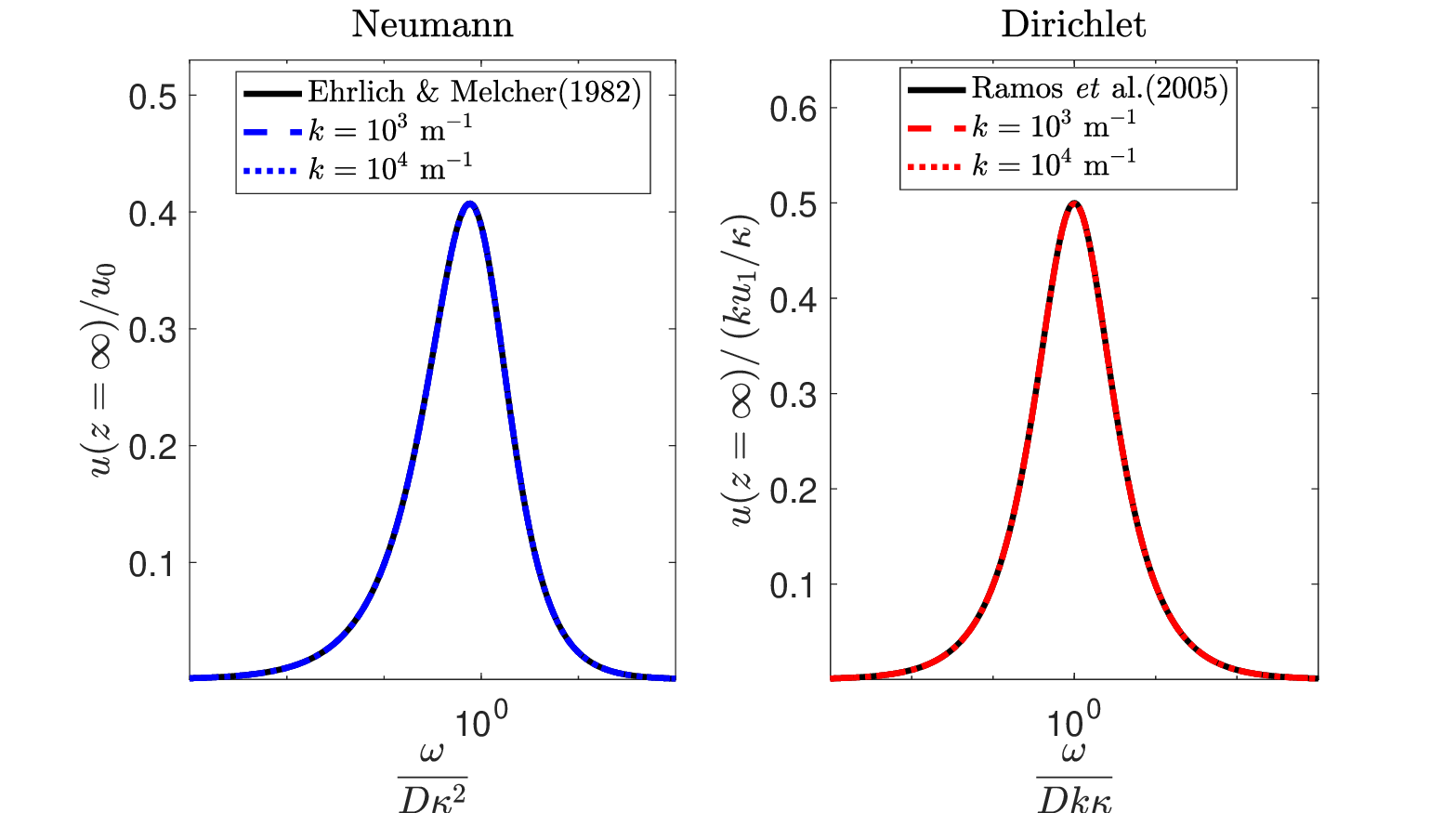}
\vspace{-0pt}
\end{center}
\caption{Horizontal \emph{zero-mode} velocity evaluated at $z= \infty$,  
from Eq. \rr{uzinf} as a function of dimensionless frequencies of the traveling wave wall charge (left panel) and as a function of the traveling electric potential at the wall (right panel), the latter was derived in Appendix \ref{sec: spacephi}. The curves are self-similar as described by relations \rr{uavssEinf} and \rr{uavssphiinf}, based however
on different dimensionless groups. Introducing the scaling that appears in the label of the vertical axis of the 
right panel, the Dirichlet results center about the
(continuous) curve introduced by  \citet[Eq. (10)]{Ramos2005}, (see
\rr{Ramos1}). The different $k$ curves in the left panel center about the \citet{ehrlich1982} expression, cf. 
Eq. \rr{EM1982}. If the value of $k$ becomes sufficiently large, say $k>10^5 \textrm{ m}^{-1}$, the amplitude of the curves
in both panels decreases. 
\label{uz_inf_E_phi2}  }
\vspace{-0pt}
\end{figure}
%----------------------------------------------------

\subsection{\label{sec: selfsimilarinf}The self-similar behavior of the slip velocity}
{The zero mode slip velocity \rr{uzpolarabs} displays a self-similar behaviour. For the dimensions of the available parameters $\kappa, k, D, u, \omega$ define a matrix of rank two. Thus, there are two dimensionless 
combinations that can formed \citep{Panton1996, Bluman1989}. The average velocity \rr{uzpolarabs} can then be expressed in the scaling form
\be \label{uavssEinf}
u(z=\infty)= \frac{1}{2}\frac{\epsilon E^2}{\eta \kappa} F\left(\frac{\omega}{D\kappa^2}, \frac{k}{\kappa}\right),
\ee
$F$ is a nonlinear function of the two independent dimensionless parameters and the front factor is the velocity scale $u_0$ \rr{u01}, as this is displayed in the vertical axis of Fig. \ref{uz_inf_E_phi2}. 

In the left panel of figure \ref{uz_inf_E_phi2} we plot the `slip' velocity \rr{uzinf}, that is, the horizontal \emph{zero-mode} velocity $u(z)$ evaluated at $z= \infty$ for various values of $k$. It is thus seen that all curves with $k$ up to say $10^5 \textrm{ m}^{-1}$ collapse into the single (continuous) curve introduced by \citet{ehrlich1982}, 
cf. Eq. \rr{EM1982}.

For comparison with results that have appeared in the more recent literature, on the right panel of figure \ref{uz_inf_E_phi2} we plot the velocity \rr{uzinf}, but for Dirichlet boundary conditions, employing the potentials \rr{phixztphi}, whose derivation was carried-out in Appendix \ref{sec: spacephi}. 
We thus display the slip velocity also in scaling form, although the dimensionless groups are now different
\be \label{uavssphiinf}
u(z=\infty) = \frac{1}{2}\frac{\epsilon k \phi_0^2}{\eta } G\left(\frac{\omega}{Dk\kappa}, \frac{k}{\kappa}\right),
\ee
$G$ is a nonlinear function of the two dimensionless variables 
and the front factor is the velocity scale $ku_1/\kappa$ \rr{u01}, as this is displayed in the label of the vertical axis of the right panel in figure \ref{uz_inf_E_phi2}. 
The curves appearing in the right panel of Fig. \ref{uz_inf_E_phi2} are thus the function $G$ where we only 
vary its first argument. Notice that, had we employed $\omega/(D\kappa^2)$ instead of $\omega/(Dk\kappa)$ to nondimensionalize 
$u(\infty)$ in \rr{uavssphiinf}, the curves in the right panel of Fig. \ref{uz_inf_E_phi2} for each different $k$
would be translated to the left and to the right of the center curve. 

All the right panel curves collapse to the slip velocity (continuous curve) determined by 
\citet[Eq. (10)]{Ramos2005}, having the form
\be \label{Ramos1}
u_{\textrm{Ramos}} = \frac{\kappa u_1}{k} \frac{\Omega}{1+ \Omega^2}, 
\ee
where $\Omega = \frac{\omega}{ Dk\kappa}$ in our notation. 

The self-similar behavior of the velocity is important as it implies that results obtained with one experimental configuration can be employed to obtain estimates of results corresponding to a different experimental configuration without the need to repeat the experiment.

\subsection{\label{sec: infscaling}Estimates of $u(\infty)$}
The preceding discussion, although informative, it does not provide an understanding of the magnitude of the 
effects, which we thus analyze in the present section.

Let us obtain an estimate of the velocity $u(\infty)$ in each case by considering the experimental values of 
\citet{Ramos2005,Garcia2006} where $k\sim 10^4 \textrm{ m}^{-1}$, voltage $\phi_0\sim 3$ V and employing standard parameter values for water $\epsilon = 7\times 10^{-10}$ F/m, $\eta = 0.001$ kg/m sec, 
$\kappa = 10^{7}\textrm{ m}^{-1}$.  

A fair comparison between the left and right panels of figure 
\ref{uz_inf_E_phi2} can be furnished by considering the dimensionless value of the velocity at the peak of the 
graph in each panel, which occurs at different frequencies. Thus, in the Dirichlet case we consider 
$\omega\sim Dk\kappa$, thus giving the peak dimensionless velocity value $\sim 0.5$ in the right panel of figure  
 \ref{uz_inf_E_phi2}, leading its dimensional 
counterpart to be equal to 
\be \label{uinfestphi}
u(\infty) = 1.57\; \textrm{cm/sec}, \quad \textrm{(Dirichlet)}
\ee
which is rather high. 
On the other hand, for the same values as above but adopting
$E =  10^5$ V/m, as in \citep{ehrlich1982}, $\omega\sim D\kappa^2$, thus giving the peak dimensionless velocity value $\sim 0.405$ in the left panel of figure  
 \ref{uz_inf_E_phi2}, leading its dimensional 
counterpart to be equal to 
\be \label{uinfest}
u(\infty) = 141 \;\mu\textrm{m/sec}, \quad \textrm{(Neumann)}
\ee
which is within the order of magnitude of experimental velocities, cf. \citep[Fig. 7]{Garcia2006} but also
\citep[Figs. 2 \& 10] {Cahill2004}. Note that the electric field varies on the length scale $1/k$, cf. \rr{phixzt}, justifying
the choice $E =  10^5$ V/m for $k\sim 10^4 \textrm{ m}^{-1}$, in determining the estimate \rr{uinfest}, for compatibility with the electric potential
value  $\phi_0\sim 3$ V employed in \rr{uinfestphi}. 

A recurrent theme in traveling wave electroosmosis studies is the high theoretical estimates (usually carried-out by employing Dirichlet boundary conditions) in comparison to much lower, by a few orders of magnitude, experimental measurements cf. \citep{Cahill2004,Ramos2005}. It is thus possible that the Neumann boundary conditions employed in this paper and their accompanying estimates, as in \rr{uinfest}, may resolve this inconsistency.

%For comparison, employing the above material parameters, we estimate the theoretical velocity determined by \citet[Eq. (10)]{Ramos2005}, having the form
%\be \label{Ramos2}
%u_{\textrm{Ramos}} = u_2 \frac{\Omega}{1+ \Omega^2} \sim 1.24\; \textrm{cm/sec}, 
%\ee
%where $\Omega = \frac{\omega}{ Dk\kappa}$ in our notation and $u_2$ is the velocity scale \rr{u012}. In \rr{Ramos2} we have set their material 
%parameter $\Lambda=1$. 

\subsection{\label{sec: infasympt}Tails of $u(\infty)$}

Both Neumann and Dirichlet tails in Fig. \ref{uz_inf_E_phi2} are of $O(\omega)$  as $\omega \rightarrow 0$:
\be
u(\infty) \sim  \frac{\kappa \omega}{4D}
\frac{k(7k^2 + 8\kappa^2) + \sqrt{k^2 + \kappa^2}(8k^2 + 4\kappa^2)}{
\left( \kappa^2 + 2 k^2 + 2k \sqrt{k^2 + \kappa^2} \right)^2 } \times
\left\{
\begin{array}{ll}
\frac{u_1}{k^2 + \kappa^2}  & \textrm{(Dirichlet)} \\
\frac{u_0\kappa^2}{(k^2 + \kappa^2)^2 }
& \textrm{(Neumann)}
\end{array}
\right. \textrm{as} \quad \omega \rightarrow 0. 
\ee

The high frequency asymptotics of the exact expressions \rr{uavssEinf} and \rr{uavssphiinf} are given by 
\be \label{highomegainf}
u(\infty) \sim  \left\{
\begin{array}{ll}
\frac{u_1}{\sqrt{2}} \left( \frac{k}{\kappa}\right)^{3/2} \left(\frac{\omega}{Dk\kappa}\right)^{-3/2}  & \textrm{(Dirichlet)} \\
 \frac{u_0}{\sqrt{2}}  \left(\frac{\omega}{D\kappa^2}\right)^{-3/2} 
& \textrm{(Neumann)}
\end{array}
\right. \textrm{as} \quad \omega \rightarrow \infty. 
\ee
Notice that the inherent frequency scalings in \rr{highomegainf} agree with the scalings of the functions
$G$ and $F$ in \rr{uavssphiinf} and \rr{uavssEinf}, respectively. 

}

\begin{figure}
\vspace{5pt}
\begin{center}
\includegraphics[height=2.2in,width=5in]{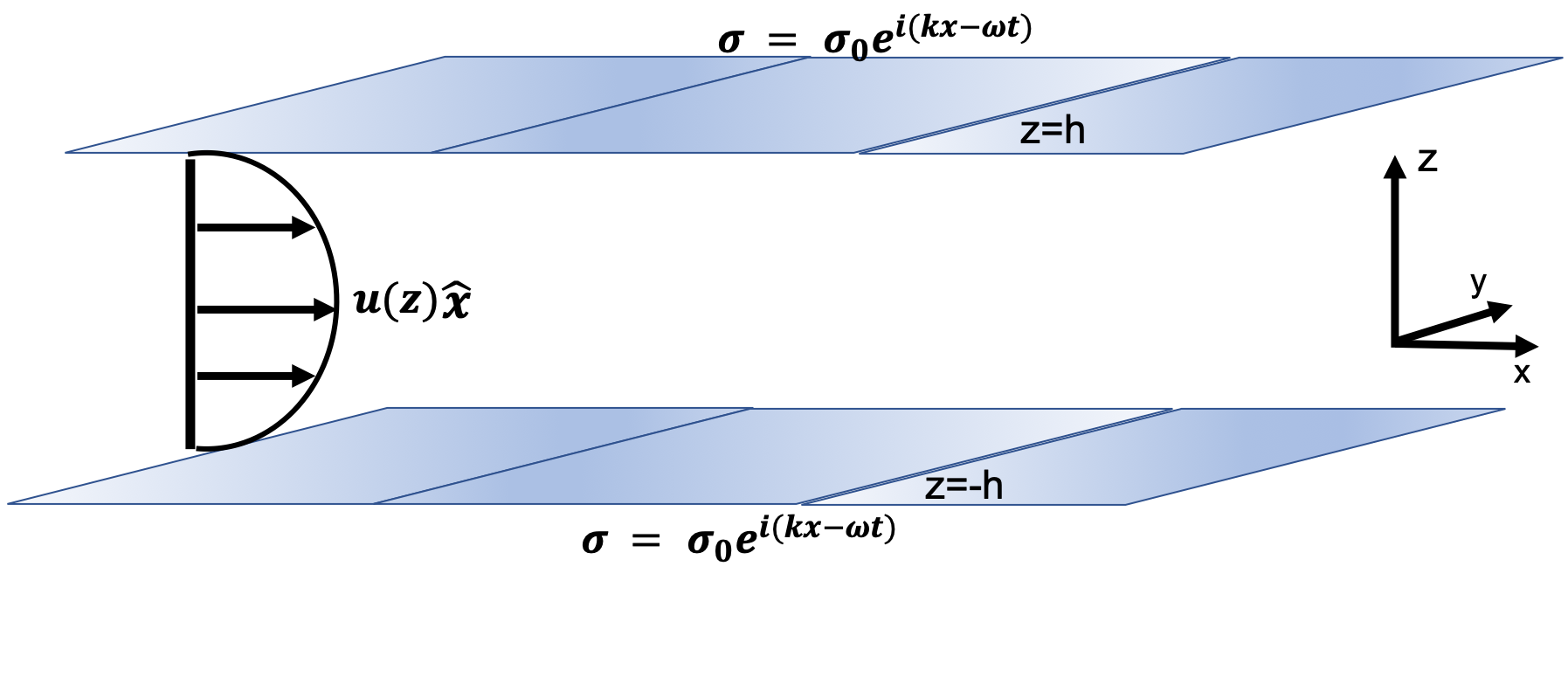}
\vspace{-0pt}
\end{center}
\caption{Traveling-wave wall charges give rise to a nonlinear body force and torque (see \rr{psinl}) in a $1:1$ electrolyte in a rectangular channel of width $2h$ leading to the appearance of a (zero mode) unidirectional fluid velocity in the $\hat{\mathbf{x}}$ direction, parallel to the channel walls that is quadratic 
with respect to the associated electric field and that does not vanish after averaging over the charge period of oscillation.
\label{channel1wave}  }
\vspace{-0pt}
\end{figure}

\section{\label{sec: channel}Traveling wave nonlinear electroosmosis in a channel of width $2h$}
Instead of the nonlinear electroosmosis taking place in a semi-infinite space as developed in section \ref{sec: space}, 
it is more physically realistic to consider the configuration displayed in Fig. \ref{channel1wave}. Traveling wave surface charges are applied on the channel walls at $z=\pm h$ enclosing
an $1:1$ electrolyte. 
The boundary conditions satisfied 
by the potential $\phi$ and by the charge $\rho$ are 
\be \label{phirhochannel}
\partial_z \phi (x,z = \pm h, t) = \pm   \frac{\sigma(x, t)}{\epsilon}, \quad \partial_z \rho (x,z = \pm h, t) =\mp \kappa^2 \sigma(x, t), 
\ee
with $\sigma= \sigma_0 e^{i (kx -\omega t)}$. 
{The choice of this set-up was dictated by results showing that the ideal and optimal configuration
seems to require symmetrical electrode arrays with respect to the channel center and with no phase lag between them
\citep{Yeh2011}. }

Assuming $\rho = \rho(z) e^{i (kx -\omega t)}$ and subject  to the 
boundary condition \rr{phirhochannel} with surface charge \rr{sigmatravel} the equations to solve are identical 
to those of the semi-infinite space. Thus, 
the charge distribution reads
\be \label{rhoxztchannel}
\rho(x,z,t) = -\frac{\sigma_0\kappa^2\cos Pz}{P\sin Ph} e^{i(kx -\omega t)} ,  
\ee
where $P$ is the  complex wavenumber defined in \rr{kdelta2}. 

%Note that this form of bulk charge distribution does not violate
%electroneutrality. The denominator of expression \rr{rhoxztchannel} is composed of hyperbolic functions of large argument (since $P$ is complex) so, close
%to the center of the channel ($z=0$) the charge is effectively zero, 
%cf. \citep[Eq. (4)]{ajdari1995} for the corresponding case of a steady periodic
%wall charge distribution. 

Similarly, assuming $\phi = \phi(z) e^{i (kx -\omega t)}$ and subject  to the 
boundary condition \rr{phirhochannel} with surface charge \rr{sigmatravel}, the potential distribution reads
\be \label{phixztchannel}
\phi(x,z,t) = \frac{\sigma_0}{\epsilon} \left( 1 + \frac{\kappa^2}{P^2 +k^2} \right)  \frac{\cosh kz}{k\sinh kh} e^{i(kx -\omega t)}
+\frac{1}{\epsilon (P^2 +k^2)} \rho, 
\ee
with $\rho$ given by \rr{rhoxztchannel}. 
{We note that, as is the case in the semi-infinite space, the above equations do not have a solution when $k=0$. For the same reasons as before, we are not interested in this limit as the physically realistic wall charge excitation wavenumbers we have in mind are very large, lying in the interval $(10^1, 10^5) \textrm{ m}^{-1}$, as this was stated in table
\ref{tab: table1}.  }
%\begin{figure}
%\vspace{5pt}
%\begin{center}
%\includegraphics[height=2.6in,width=3.8in]{uomega2}
%\vspace{-0pt}
%\end{center}
%\caption{
%Zero mode velocity \rr{uavchannel}, averaged over the channel width $2h$.
%Following 
%\citep{ehrlich1982}, the velocity has been scaled by $u_0 = \frac{1}{2}\frac{\epsilon E^2}{\eta \kappa}$, 
%and frequency by $D\kappa^2$, where $E = \frac{\sigma_0}{\epsilon}$ is the amplitude of the induced electric field due to the
%interfacial charge distribution $\sigma_0$, $\kappa^{-1}, D, \eta$ are the Debye length, charge diffusion coefficient
%and liquid viscosity, respectively. Here we have taken $k =10^3 \textrm{ cm}^{-1}$, $\kappa =10^4 \textrm{ cm}^{-1}$
%$D = 10^{-5}\textrm{ cm}^{2}/$sec, $h=10^{-3}$ cm and $\omega = D\kappa^2$. 
%\label{uomega}  }
%\vspace{-0pt}
%\end{figure}
%

The zero mode velocity in the channel again satisfies the integrated momentum Eq. \rr{upz}. 
Employing the complex form of the field $\phi$ in \rr{phixztchannel} we thus solve \rr{upz} subject to the no-slip boundary conditions
\be \label{upzbcchannel}
u(z=\pm h)=0.
\ee
The zero mode velocity $u(z)$ from \rr{upz} with \rr{phixztchannel} becomes 
\be \label{uzchannel}
u(z) = \frac{i\epsilon k}{4\eta} \int_h^z \left( \phi^*\phi_z - \phi \phi_z^*\right) dz
\ee
Following the same steps as in the semi-infinite space of section \ref{sec: space}, leads to a long and uninformative expression for $u(z)$. This is 
essentially a plug-like flow with very thin boundary layers of thickness $\sim  \kappa^{-1}$ located at $z =\pm h$. 
The meaningful observable here is the average value of the zero mode over the width of the channel
\be \label{uavchannel}
\langle  u \rangle = \frac{1}{2h} \int_{-h}^h {u(z)} dz. 
\ee
%Following 
%\citet{ehrlich1982}, we will scale the velocity by $u_0 = \frac{1}{2}\frac{\epsilon E^2}{\eta \kappa}$, 
%where $E = \frac{\sigma_0}{\epsilon}$ is the amplitude of the electric field due to the
%interfacial charge distribution $\sigma_0$. 
As it stands, \rr{uavchannel} is a double integral and could be awkward  to evaluate, especially
in cylindrical polars (see section \ref{sec: cylinder}). It can be significantly simplified by changing the order 
of integration and carrying-out one of the integrations. It can thus be written as a single integral in the form
\be \label{uavchannel2}
\langle  u \rangle = -\frac{1}{2h} \frac{i\epsilon k}{4\eta}\int_{-h}^h \left( \phi^*\phi_z - \phi \phi_z^*\right)  zdz,
\ee
where the parity of $\phi$ with respect to the origin $z=0$ was taken into account.

%-----------------------------------------------------
% figure(1) in electroosmosis_wave3_channel_Ephi.m
%----------------------------------------------------
\begin{figure}
\vspace{5pt}
\begin{center}
\includegraphics[height=2.6in,width=5.6in]{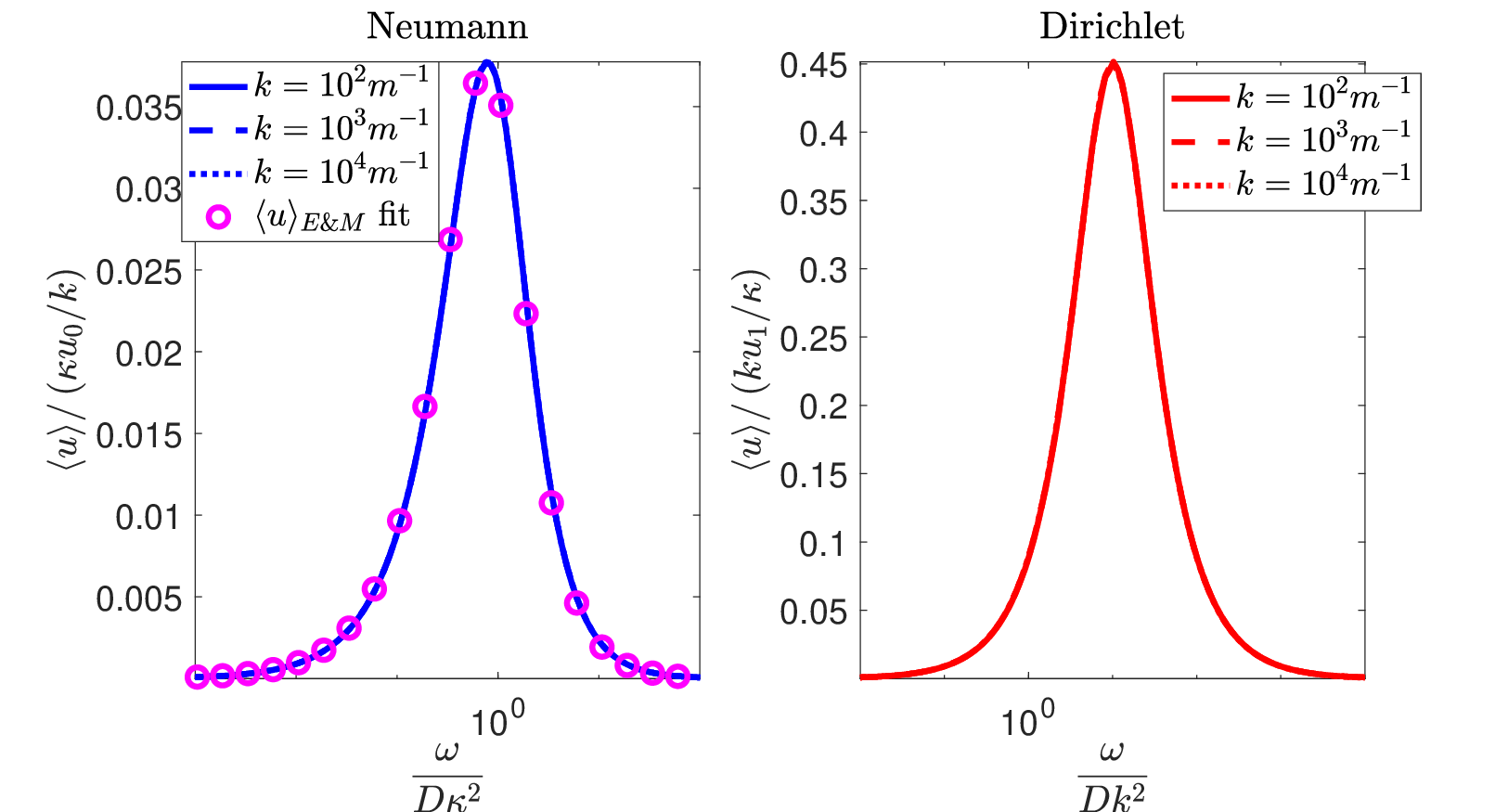}
\vspace{-0pt}
\end{center}
\caption{Self-similar behaviour of the 
zero mode velocity \rr{uavchannel}, averaged over the channel width $2h$
as a function of dimensionless frequencies of the traveling wave wall charge (left panel) or of the traveling electric potential at the wall (right panel), the latter was derived in Appendix \ref{sec: spacephi}. The curves are self-similar as described by relations \rr{uavssEchannel} and \rr{uavssphichannel}. We emphasize that the dimensionless
group $\omega/(Dk^2)$ employed in the right panel, \emph{differs} from the group $\omega/(D\kappa^2)$ 
employed in the left panel, and furthermore, it is different to the group $\omega/(Dk\kappa)$ employed for the Dirichlet case in the 
semiinfinite space (right panel of Fig. \ref{uz_inf_E_phi2}).
Both panels: $h = 10^{-5}$ m, $D = 10^{-9}\textrm{ m}^{2}/$sec and in the left panel we employed the fit \rr{EM1982b} with the 
single parameter $\beta = 0.926$. 
%If the value of $k$ becomes sufficiently large, say $k>10^5 \textrm{ m}^{-1}$, the amplitude of the curves
%in both panels decreases. 
\label{uav_selfsimilar_phi_channel}  }
\vspace{-0pt}
\end{figure}

%
%\begin{figure}
%\vspace{5pt}
%\begin{center}
%\includegraphics[height=2in,width=5.6in]{uav_selfsimilar_E_channel}
%\vspace{-0pt}
%\end{center}
%\caption{Self-similar behaviour of the 
%zero mode velocity \rr{uavchannel2}, averaged over the channel width $2h$. 
%It is expressed with respect to the dimensionless groups in \rr{uavssEchannel}. Circles denote the theoretical fit expression 
%\rr{EM1982b}, adopted from \citep{ehrlich1982} with an appropriate rescaling. 
%$D = 10^{-9}\textrm{ m}^{2}/$sec in both panels.
%\label{uav_selfsimilar_E_channel}  }
%\vspace{-0pt}
%\end{figure}

{
\subsection{\label{sec: selfsimilarchannel}The self-similar behaviour of the average velocity in the channel}
The zero mode velocity \rr{uavchannel2} displays a self-similar behaviour. For the dimensions of the available parameters $h, \kappa, k, D, u, \omega$ define a matrix of rank two. Thus, there are two dimensionless 
combinations that can formed \citep{Panton1996, Bluman1989}. The average velocity \rr{uavchannel} can be given in the form
\be \label{uavssEchannel}
\langle  u \rangle = \frac{1}{2}\frac{\epsilon E^2}{\eta k} f\!\left(\frac{\omega}{D\kappa^2}, \kappa h\right),
\ee
$f$ is a nonlinear function of the two independent dimensionless parameters and the front factor is the velocity scale $\kappa u_0/k$, as this is displayed in the label of the vertical axis of Fig. \ref{uav_selfsimilar_phi_channel}
(the scaling theory allows other combinations as well, for instance see Eq. \rr{uavasom} below). 
The front factor in \rr{uavssEchannel} implies that $\langle  u \rangle$ scales as $k^{-1}$. This behavior is 
clearly displayed by the continuous curve in Fig. \ref{uavk} of Appendix \ref{sec: kscaling}. Such a behavior
for Neumann boundary conditions was met before \citep[Fig. 6(f)]{Liu2018}. Therein, increasing the 
size of the system gave rise to a commensurate increase of the Coulomb force and the resulting slip velocity. 
The curve in figure 6(f) of this reference has exactly slope equal to $1$, that is, the slip velocity scales linearly
with respect to system size, and is thus inversely proportional to $k$, as this is 
demonstrated by the front factor in \rr{uavssEchannel} and the continuous curve in Fig. \ref{uavk}.

A good single-parameter fit of these results that accounts for all parameter values, including the tails, is given by the circle points in Fig. \ref{uav_selfsimilar_phi_channel}, 
and has the functional form 
\be \label{EM1982b}
\langle  u \rangle_{E\&M} = \frac{\beta}{ \kappa h} \frac{\frac{\omega}{D\kappa^2} }{\left[\left(\frac{\omega}{D\kappa^2}\right)^{2}+1\right]^{\frac{5}{4}}}\cos \! \left(\frac{\arctan \left(\frac{\omega}{D\kappa^2}\right)}{2}\right),
\ee
i.e. it is the velocity \rr{EM1982} scaled accordingly and the single parameter value $\beta = 0.926$.

Thus, experimental results obtained by employing a single configuration that agree with the above behaviour can be
extrapolated theoretically to fit alternative experimental conditions and parameters without the need of repeating the experiments.  

%In Appendix \ref{sec: selfsimilarchannelphi} we repeat the above discussion but for the case of Dirichlet boundary conditions for the Poisson-Nernst-Planck equations. Likewise, one obtains a self-similar behavior of the averaged zero mode velocity \rr{uavchannel}, although for different dimensionless groups, cf. Fig. \ref{uav_selfsimilar_phi_channel},

Employing the Dirchlet boundary conditions instead, we obtain the potential \rr{phixztchannelphi} and thus the velocity \rr{uavchannel2} acquires the self-similar form
\be \label{uavssphichannel}
\langle  u \rangle = \frac{1}{2}\frac{\epsilon k \phi_0^2}{\eta } g\!\left(\frac{\omega}{Dk^2}, \kappa h\right),
\ee
$g$ is a nonlinear function of the two independent dimensionless parameters and the front factor is the velocity scale $k u_1/\kappa$, as this is displayed in the vertical axis of the right panel in Fig. \ref{uav_selfsimilar_phi_channel}. Notice however that the dimensionless
group $\omega/(Dk^2)$ employed in \rr{uavssphichannel} (right panel of Fig. \ref{uav_selfsimilar_phi_channel}), \emph{differs} from the group $\omega/(D\kappa^2)$ 
employed in \rr{uavssEchannel} (left panel of Fig. \ref{uav_selfsimilar_phi_channel}) and furthermore, it is different to the group $\omega/(Dk\kappa)$ employed for the Dirichlet case in the 
semiinfinite space in \rr{uavssphiinf} (right panel of Fig. \ref{uz_inf_E_phi2}).

%-----------------------------------------------------
% figure(9) in electroosmosis_wave3_channel_Ephi.m
%----------------------------------------------------
\begin{figure}
\vspace{5pt}
\begin{center}
\includegraphics[height=2.4in,width=5.6in]{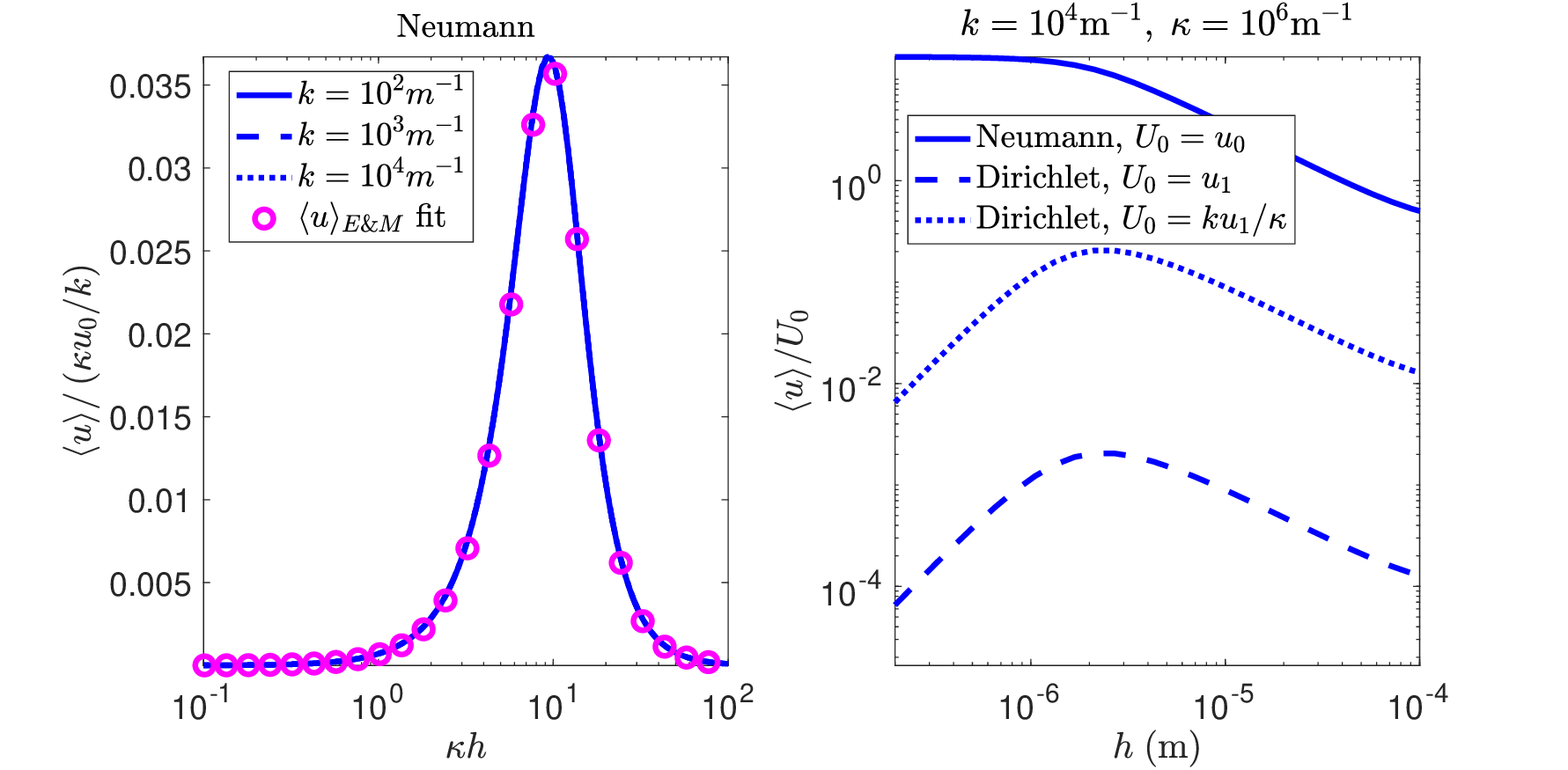}
\vspace{-0pt}
\end{center}
\caption{Left panel: Self-similar behavior of average velocity by varying only the scaled channel width
(second argument of  \rr{uavssEchannel}). Circles denote the fit \rr{EM1982b}. Right panel: 
Averaged over the channel width zero mode velocity \rr{uavchannel2} vs. channel height $h$, obtained with Neumann or Dirichlet boundary conditions (employing the potentials \rr{phixztchannel} and \rr{phixztchannelphi}, respectively) and scaled by either $u_0$, $u_1$ or $ku_1/\kappa$ (cf. \rr{u01}).  The Neumann velocity reaches a plateau
for small channel heights while its Dirichlet counterpart decays to zero. For larger values of $h$ all curves reach
plateaus. Curves obtained with a frequency corresponding to the peak velocity of the left and the right panels
of Fig.  \rr{uav_selfsimilar_phi_channel}, respectively. 
\label{uav_h_Ephi}  }
\vspace{-0pt}
\end{figure}
%------------------------------------------------------------

\subsection{\label{sec: channelscaling}Average velocity $\langle u \rangle$ estimates}
%Figure \ref{uav_k_Ephi} displays the $k$-dependence of the averaged over the channel width zero mode velocity \rr{uavchannel2} obtained with Neumann or Dirichlet boundary conditions (employing the potentials \rr{phixztchannel} and \rr{phixztchannelphi}, respectively) and scaled by either $u_0$, $u_1$ or $u_2$ (cf. \rr{u012}).  Each line, from top to bottom, has slope $-1, 2$ and $3$ respectively, determining the velocity $k-$ dependence as $k^{-1}, k^2$ and $k^3$, respectively. 

We estimate the channel average velocity by considering the same material parameters as in 
subsection \ref{sec: infscaling}. 
Comparison between the left and right panels of figure 
\ref{uav_selfsimilar_phi_channel}  can be furnished by considering the dimensionless value of the velocity at the peak of the 
graph in each panel, which occurs at different frequencies. In the Dirichlet case we consider 
$\omega\sim Dk^2$, thus giving the peak dimensionless velocity value $\sim 0.45$ in the right panel of figure  
 \ref{uav_selfsimilar_phi_channel}, leading its dimensional 
counterpart to equal  
\be
\langle u \rangle = 1.4\; \textrm{cm/sec}, \quad \textrm{(Dirichlet)}.
\ee
On the other hand, for the same values as above but adopting
$E =  10^5$ V/m, as in \citep{ehrlich1982}, $\omega\sim D\kappa^2$, thus giving the peak dimensionless velocity value $\sim 0.037$ in the left panel of figure  
 \ref{uav_selfsimilar_phi_channel}, leading its dimensional 
counterpart to equal 
\be
\langle u \rangle  = 1.3 \textrm{ cm/sec}, \quad \textrm{(Neumann)} 
\ee
which are comparable to each other. Notice however, that the velocity scaling of the Dirichlet problem
$k u_1/\kappa$ decreases linearly with increasing $k$, and that the Neumann scaling $\kappa u_0/k$, increases, cf. labels of the vertical axes of the left and right panels in figure 
\ref{uav_selfsimilar_phi_channel}, respectively.

\subsection{\label{sec: channelasympt}Average velocity $\langle u \rangle$ asymptotic behavior}
In figure \ref{uav_h_Ephi} we display the characteristic behavior of the average velocity \rr{uavchannel2} 
as the height $h$ of the channel varies. The Neumann velocity in general increases as $h$ decreases
reaching a plateau at small channel heights (where the Debye layers from adjacent walls start to overlap). 
The Dirichlet velocity decays to zero in this limit. These behaviors are described by the limiting expressions
\be \label{uavh0}
\langle u \rangle \sim \left\{
\begin{array}{ll}
\frac{k u_1}{\kappa}
\frac{\frac{\omega}{Dk^2}}{3\left[ 1 +  \left( \frac{\omega}{Dk^2} \right)^2 \right]} \left( \kappa h\right)^2 & \textrm{(Dirichlet)} \\
\frac{\kappa  u_0}{k}\frac{2}{3} \frac{-\kappa^2 P_1P_2 \left[ P_1^2 +(P_2+k)^2 \right]^3\left[ P_1^2 +(P_2-k)^2 \right]^3 }{|P|^4\left[ P^2+k^2 \right]^3\left[ (P^*)^2+k^2 \right]^3}
& \textrm{(Neumann)}
\end{array}
\right. \textrm{as} \quad h\rightarrow 0,
\ee
where $P$ is defined in \rr{kdelta2} and $u_0$ and $u_1$ in \rr{u01}. Combination of this result 
with the $k$ scaling behavior displayed in Fig. \ref{u_k_loglog}, leads to the conclusion that 
the Neumann conditions provide prominent average velocities in thin and long channels (where 
the wavelength of the charge excitation at the walls can become sufficiently long). 

Both Neumann and Dirichlet tails in Fig. \ref{uav_selfsimilar_phi_channel}  are of $O(\omega)$ (whose 
coefficients are too long to include here), as $\omega \rightarrow 0$.
The high frequency asymptotics of the exact expressions \rr{uavssEchannel} and \rr{uavssphichannel} are given by 
\be \label{uavasom}
\langle u \rangle \sim  \left\{
\begin{array}{ll}
u_1 \frac{1}{\sqrt{2}}  \left(\frac{\omega}{Dk^2}\right)^{-3/2} \tanh kh & \textrm{(Dirichlet)} \\
 \frac{u_0}{\sqrt{2}\tanh kh}  \left(\frac{\omega}{D\kappa^2}\right)^{-3/2} 
& \textrm{(Neumann)}
\end{array}
\right. \textrm{as} \quad \omega \rightarrow \infty. 
\ee
It is clear that the expressions in \rr{uavasom} justify the frequency scalings of the functions $g$ and $f$ 
in \rr{uavssphichannel} and \rr{uavssEchannel}, respectively. 

%
%$\langle  u \rangle =\frac{1}{2}\frac{\epsilon \kappa  \phi_0^2}{\eta } \frac{1}{\sqrt{2}}  \left(\frac{\omega}{Dk^2}\right)^{-3/2} \tanh kh + O(\omega^{-5/2})$. 
%
%
%
%The high frequency asymptotics of the exact expression \rr{uavssEchannel} are given by 
%$\langle  u \rangle =\frac{1}{2}\frac{\epsilon E^2}{\eta \kappa} \frac{1}{\sqrt{2}\tanh kh}  \left(\frac{\omega}{D\kappa^2}\right)^{-3/2}  + O(\omega^{-5/2})$. 

}

\section{\label{sec: cylinder}Traveling wave nonlinear electroosmosis in a cylindrical capillary}
A still more physically realistic configuration is displayed in Fig. \ref{channel1wave}. Traveling wave surface charges 
$\sigma = \sigma_0 e^{i(kz - \omega t)}$
are applied to the wall of a capillary with circular cross-section of radius $a$,  enclosing
an $1:1$ electrolyte. 
The boundary conditions satisfied 
by the potential $\phi$ and by the charge $\rho$ are 
\be \label{phirhocylinder}
\partial_r \phi (z,r =a, t) =  \frac{\sigma(z, t)}{\epsilon}, \quad \partial_r \rho (z,r=a, t) = - \kappa^2 \sigma(z, t), 
\ee
\begin{figure}
\vspace{5pt}
\begin{center}
\includegraphics[height=1.6in,width=3.8in]{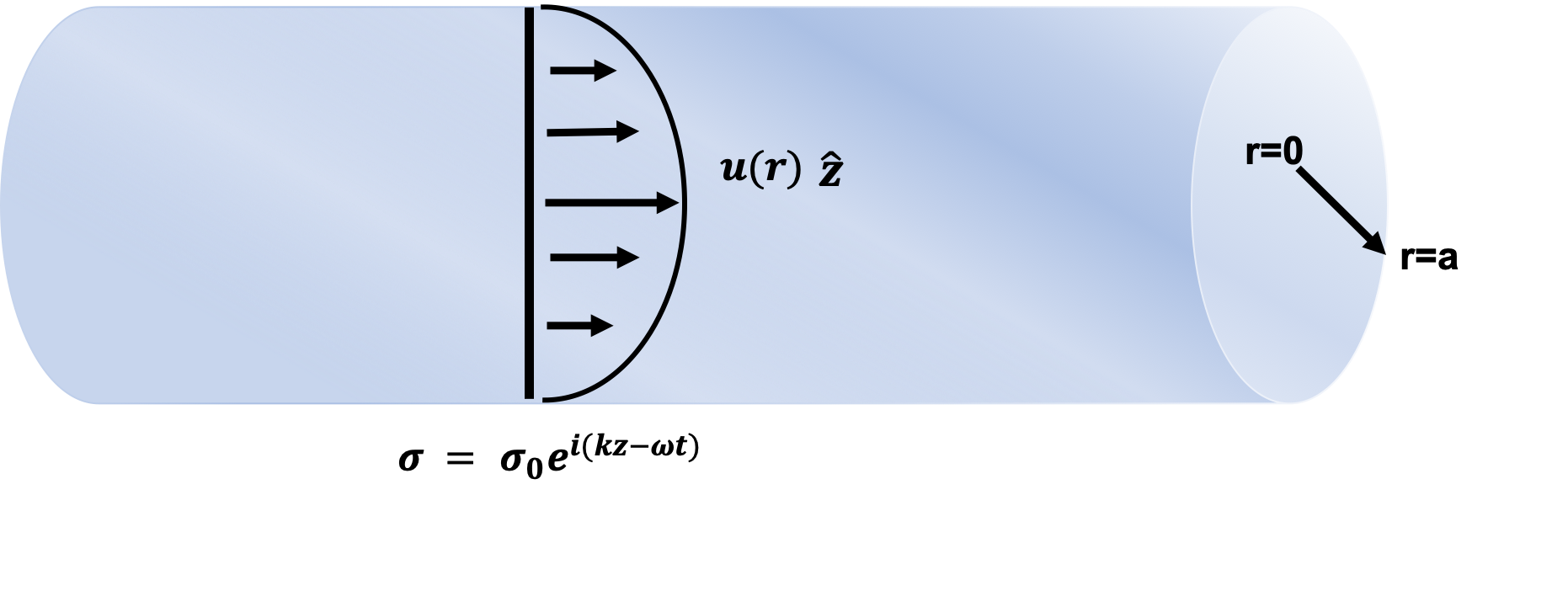}
\vspace{-0pt}
\end{center}
\caption{Traveling-wave wall charges give rise to a nonlinear body force and torque (see \rr{psinl}) in a $1:1$ electrolyte in a cylindrical capillary of radius $a$ leading to the appearance of a unidirectional fluid velocity in the $\hat{\mathbf{z}}$ direction, along the capillary center axis, that is quadratic 
with respect to the associated electric field and that does not vanish after averaging over the charge period of oscillation.
\label{cylinder1wave}  }
\vspace{-0pt}
\end{figure}
and we take the axis of the cylinder to lie in the $x$-direction as displayed in Fig. \ref{cylinder1wave}. 
Assuming $\rho = \rho(r) e^{i (kz -\omega t)}$ and $\phi = \phi(r) e^{i (kz -\omega t)}$, the evolution equation for the charge and Gauss law reduce to
\be \label{rhophicylinder}
\rho_{rr} + \frac{1}{r} \rho_r + \left[ \frac{i\omega}{D} - k^2 - \kappa^2 \right] \rho = 0, \quad \textrm{and}  \quad 
\phi_{rr} + \frac{1}{r} \phi_r  - k^2 \phi = - \frac{\rho}{\epsilon}, 
\ee
respectively. 
With 
boundary conditions \rr{phirhocylinder}, we obtain
\be \label{rhoxztcylinder}
\rho(z,r,t) = -\frac{\sigma_0\kappa^2J_0( Pr)}{PJ_1( Pa)} e^{i(kz -\omega t)} ,  
\ee
where $P$ is (again) the  complex wavenumber \rr{kdelta2} and $J_0$, $J_1$ are Bessel functions of the first kind. 
%As in the case of the channel, Eq. \rr{rhoxztcylinder} does not violate electroneutrality, see the discussion below \rr{rhoxztchannel}.
Likewise, subject  to the 
boundary condition \rr{phirhocylinder}, the potential distribution reads
\be \label{phixztcylinder}
\phi(z,r,t) = \frac{\sigma_0}{\epsilon} \left( 1 + \frac{\kappa^2}{P^2 +k^2} \right)  \frac{I_0(kr)}{k I_1( ka)} e^{i(kz -\omega t)}
+\frac{1}{\epsilon (P^2 +k^2)} \rho, 
\ee
with $\rho$ given by \rr{rhoxztcylinder} and $I_0$ and $I_1$ are modified Bessel functions of the first kind (Bessel functions of imaginary argument).

\begin{figure}
\vspace{5pt}
\begin{center}
\includegraphics[height=2.4in,width=3.4in]{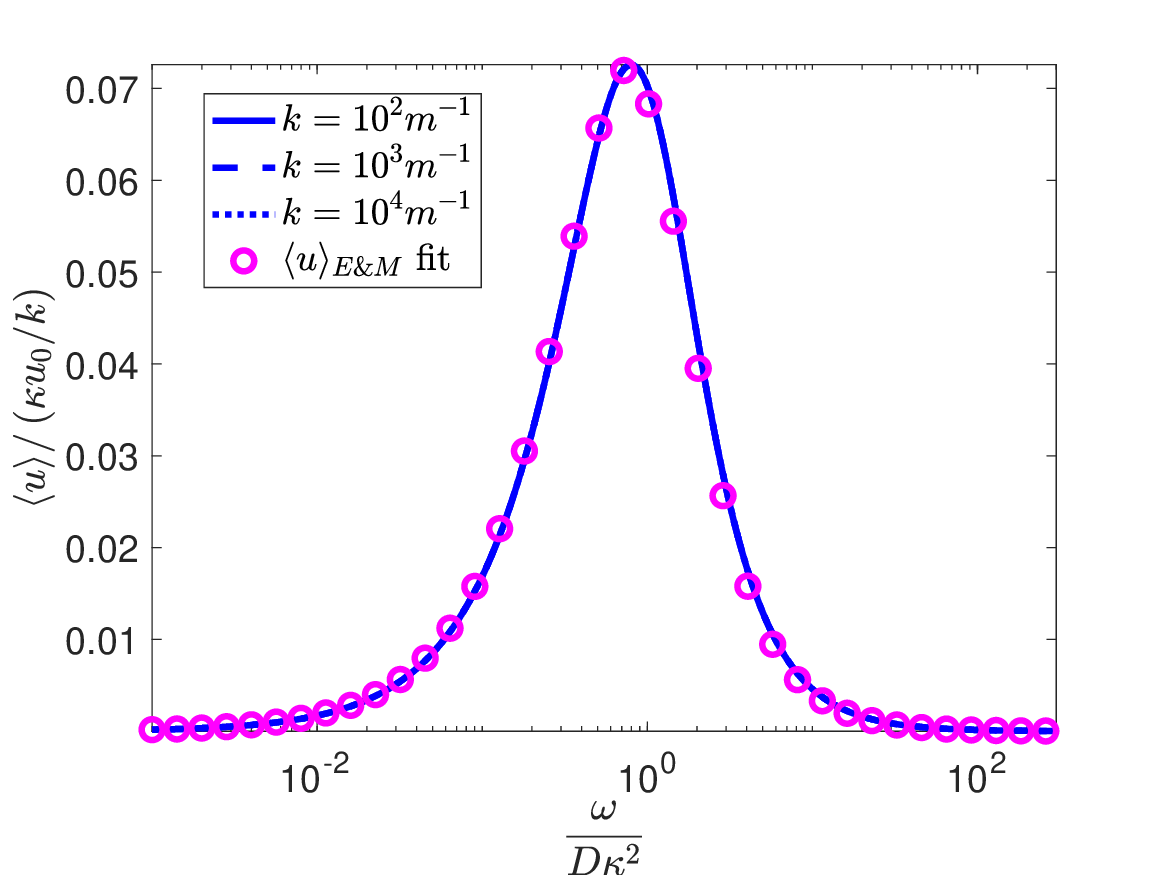}
\vspace{-0pt}
\end{center}
\caption{Traveling-wave electroosmosis in a cylinder, cf. Fig. \ref{cylinder1wave}, with traveling wave charge distribution (that is, Neumann boundary conditions for the Poisson equation). 
Self-similar behaviour of the 
zero mode velocity \rr{uavcylinder}, averaged over the cross-sectional area of the cylinder of radius $a$
as a function of the dimensionless frequency $\omega/(D\kappa^2)$ of the traveling wave wall charge. The curves are self-similar as described by relation \rr{uavssEcylinder}. The circle 
markers denote the fit \rr{EM1982c} by adopting the single parameter value $\beta = 1.77$. 
Both panels: $h = 10^{-5}$ m, $D = 10^{-9}\textrm{ m}^{2}/$sec. 
%If the value of $k$ becomes sufficiently large, say $k>10^5 \textrm{ m}^{-1}$, the amplitude of the curves
%decreases. 
\label{uomegacylinder}  }
\vspace{-0pt}
\end{figure}

As before, the velocity field consists of terms that vanish after averaging over the period of oscillation of the fields.
The exception is the zero-mode velocity, satisfying an equation analogous to \rr{upz}.
Starting from the time-dependent Stokes equations
and considering the velocity field to have the form $\mathbf{v} = u(r) \hat{\mathbf{z}}$ (as depicted in Fig. \ref{cylinder1wave}), we obtain
\be \label{urr}
u_{rr} + \frac{1}{r} u_r + \frac{ik}{4\eta} (\rho \phi^* - \rho^* \phi)  =0. 
\ee
Using the second of \rr{rhophicylinder} and performing one integration we arrive at 
\be \label{upr}
\partial_ru(r) = \frac{i\epsilon k}{4\eta}  \left( \phi^*\phi_r - \phi \phi_r^*\right)
\ee
which is the cylindrical counterpart of \rr{upz}. 
Employing the complex form of the field $\phi$ in \rr{phixztcylinder} we thus solve \rr{upr} subject to the no-slip boundary condition
\be \label{upzbccylinder}
u(r=a)=0. 
\ee
The zero mode velocity $u(r)$ from \rr{upr} with \rr{phixztcylinder} becomes symbolically 
\be \label{urcylinder}
u(r) = \frac{i\epsilon k}{4\eta} \int_a^r \left( \phi^*\phi_r - \phi \phi_r^*\right)dr,
\ee
which is the cylindrical counterpart of \rr{uzinf} and \rr{uzchannel} (the constant of integration is zero to maintain $\partial_r u |_{r=0} =0$.
Otherwise the flow would have a cusp (discontinuous derivative of $u$) at $r=0$).  
The same steps followed in the semi-infinite space problem of section \ref{sec: space}, lead to a long expression for $u(r)$ which is a plug-like flow with very thin boundary layers of thickness $\sim  \kappa^{-1}$ located at the cylinder wall $r=a$. 
The meaningful observable here is the average value of the zero mode over the capillary cross-sectional area, defined as
\be \label{uavcylinder}
\langle  u \rangle = \frac{1}{\pi a^2} \int_{0}^a {u(r)} 2\pi r dr, 
\ee
where $u(r)$ is given by \rr{urcylinder}. Eq. \rr{uavcylinder} is a double integral and involves integrals of Bessel function pairs which,  in general, 
cannot be calculated in closed form (for exceptions, see for instance %\citep{Erdelyi1953, Erdelyi1955, 
\citet{Luke1962}). We circumvent this complication by exchanging the order of integration in the double integral appearing in Eq. \rr{uavcylinder}. Thus, \rr{uavcylinder} can be written as a single integral
\be \label{uavcylinder2}
\langle  u \rangle = \frac{-1}{a^2} \frac{i\epsilon k}{4\eta}  \int_{0}^a \left( \phi^*\phi_r - \phi \phi_r^*\right) r^2 dr, 
\ee
and is calculated numerically by simple quadrature.
In figure \ref{uomegacylinder} we display the zero mode velocity averaged over the channel width \rr{uavcylinder2} versus the 
frequency of charge oscillation $\omega$ scaled by the frequency $D\kappa^2$, where $D$ is the diffusion coefficient
of the charge distribution.

{

\subsection{\label{sec: selfsimilarcylinder}The self-similar behaviour of the average velocity in the cylindrical capillary}
The zero mode velocity \rr{uavcylinder2} displays a self-similar behaviour for the same reason discussed in the 
channel case. Here the available parameters are $a, \kappa, k, D, u, \omega$ and again their dimensions define a matrix of rank two. Thus, there are two dimensionless 
combinations that can formed \citep{Panton1996, Bluman1989}. The average velocity \rr{uavcylinder2} can be given in the form
\be \label{uavssEcylinder}
\langle  u \rangle = \frac{1}{2}\frac{\epsilon E^2}{\eta k} \mathcal{F}\!\left(\frac{\omega}{D\kappa^2}, \kappa a\right),
\ee
$\mathcal{F}$ is a nonlinear function of the two independent dimensionless parameters and the front factor is the velocity scale $\kappa u_0/k$, as this is displayed in the label of the vertical axis of Fig. \ref{uomegacylinder}.
A good single parameter fit of these results that accounts for all parameter values, including the tails in Fig. \ref{uomegacylinder}, 
has the functional form 
\be \label{EM1982c}
\langle  u \rangle_{E\&M} = \frac{\beta}{ \kappa a} \frac{\frac{\omega}{D\kappa^2} }{\left[\left(\frac{\omega}{D\kappa^2}\right)^{2}+1\right]^{\frac{5}{4}}}\cos \! \left(\frac{\arctan \left(\frac{\omega}{D\kappa^2}\right)}{2}\right),
\ee
i.e. it is the velocity \rr{EM1982} scaled accordingly and we have adopted the single parameter value $\beta = 1.77$
in Fig. \ref{uomegacylinder}.

In Fig. \ref{uomegacylinder} we plot the function $\mathcal{F}$ from \rr{uavssEcylinder} versus the dimensionless
(Debye) frequency for a number of values of the excitation wavenumber $k$. The circle 
markers denote the fit \rr{EM1982c}. This form of behavior could thus be employed to reproduce results of an
experiment without the need to repeat it.

We estimate the cylinder average velocity by considering the same material parameters as in 
subsection \ref{sec: infscaling}. 
Adopting
$E =  10^5$ V/m, as in \citep{ehrlich1982}, $\omega\sim D\kappa^2$, thus giving the peak dimensionless velocity value $\sim 0.0715$ in figure  
 \ref{uomegacylinder}, leading its dimensional 
counterpart to equal 
\be
\langle u \rangle  = 2.5 \textrm{ cm/sec}. %, \quad \textrm{(Neumann)} 
\ee
Notice however, that this estimate decreases with increasing $k$.

\section{\label{sec: numerical}Numerical results}
In this section we solve the Poisson-Nernst-Planck-Navier-Stokes differential equations
numerically with a finite-element package (Comsol). 
We are interested in establishing the existence of the zero mode velocity and how does it compare
in magnitude and its general trend with the exact solutions obtained in the previous section. 

We employ the cylindrical capillary geometry of section \ref{sec: cylinder} amended so as to incorporate 
effects of boundaries in the longitudinal direction of the cylinder 
as displayed in Figure \ref{cylinder} and explained in more detail in the Supplementary Materials addendum. 
We employ the low P\'eclet number approximation to the Poisson-Nernst-Planck-Navier-Stokes differential equations
which, for the material parameters employed in this paper, is a valid approximation as shown in Appendices 
\ref{sec: linearization} and \ref{sec: Pe}. 

The equations employed are the same as previously stated, with the exception of 
the full evolution of charge distribution %(due to all three aforementioned processes),
\be  \label{rhotgeneral2}
\partial_t \rho = D\left[ \nabla^2 \rho + \frac{e}{k_BT} \nabla \cdot \left( s \nabla \phi \right) \right],
\ee
and the time-dependent Stokes equations
\be \label{NS}
\rho_l\frac{\partial \mathbf{v}}{\partial t} = -\nabla p + \eta \nabla^2 \mathbf{v} + \rho \nabla \phi, \quad \nabla \cdot \mathbf{v} = 0. 
\ee

\begin{figure}
\vspace{5pt}
\begin{center}
\includegraphics[height=1.5in,width=5.2in]{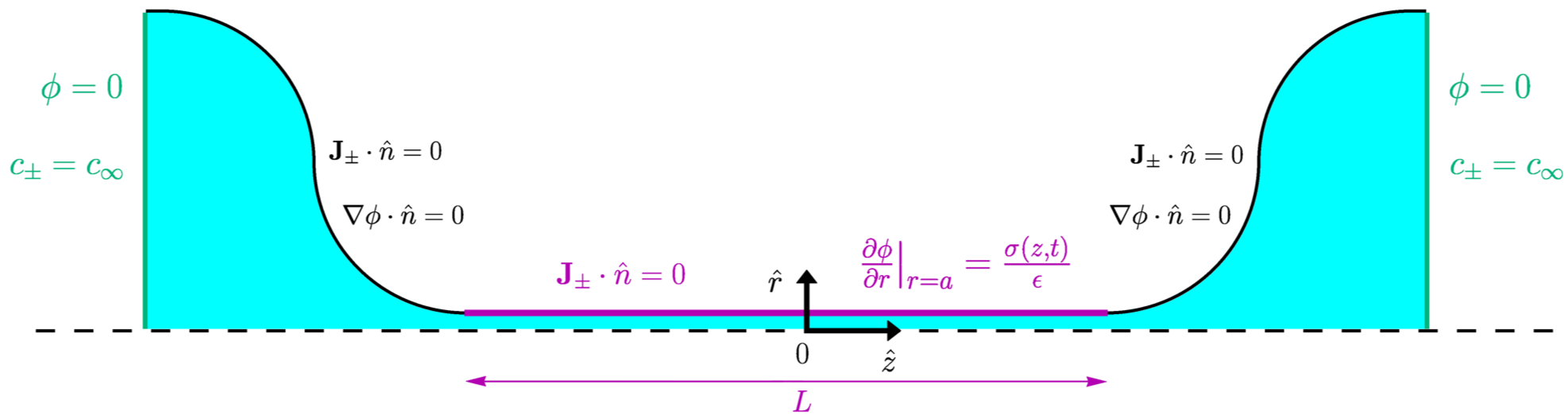}
\vspace{-0pt}
\end{center}
\caption{Configuration of the domain for the numerical solution of the Poisson-Nernst-Planck-Navier-Stokes
system in section \ref{sec: numerical}. 
\label{cylinder}  }
\vspace{-0pt}
\end{figure}

The set-up of the system is shown in Figure \ref{cylinder}. The cylinder has length $L$ and is connected to 
two reservoirs of fixed charge concentration and zero electric potential (left-most and right-most 
edges of the figure). The boundary conditions in the lateral surface 
of the cylinder are the same as those employed in section 5 of the manuscript and we employ the real part of the wall
charge density (thus, $\sigma(z,t) = \sigma_0 \cos(kz - \omega t)$). The intermediate regions
connecting the two reservoirs to the cylinder have  boundary conditions of zero normal electric
field and zero normal current $\hat{\mathbf{n}}\cdot \mathbf{J}_\pm$, where 
\be
\mathbf{J}_\pm = -D\left[ \nabla c_\pm \pm \frac{e}{k_BT}  c_\pm \nabla \phi \right]
\ee
and $\hat{\mathbf{n}}$ is the normal vector to the interface. 
We employ the parameters 
\begin{align} \nonumber
&a = 10^{-5} \textrm{ m}, \quad L =2\times 10^{-4} \textrm{ m}\textrm{ or } L = 4\times 10^{-4} \textrm{ m}, \quad D = 10^{-9} \textrm{ m}^2/\textrm{sec},  \\
&
\sigma_0 = 7\times 10^{-5} \textrm{ C/m}^2,\kappa = 2.3 \times 10^6 \textrm{ m}^{-1}, k = 0.8 \times 10^5 \textrm{ m}^{-1}, 2c_\infty = 10^{-3} \textrm{ mM}.  
\label{paramN}
\end{align}

\begin{figure}
\vspace{5pt}
\begin{center}
\includegraphics[height=2.3in,width=5.8in]{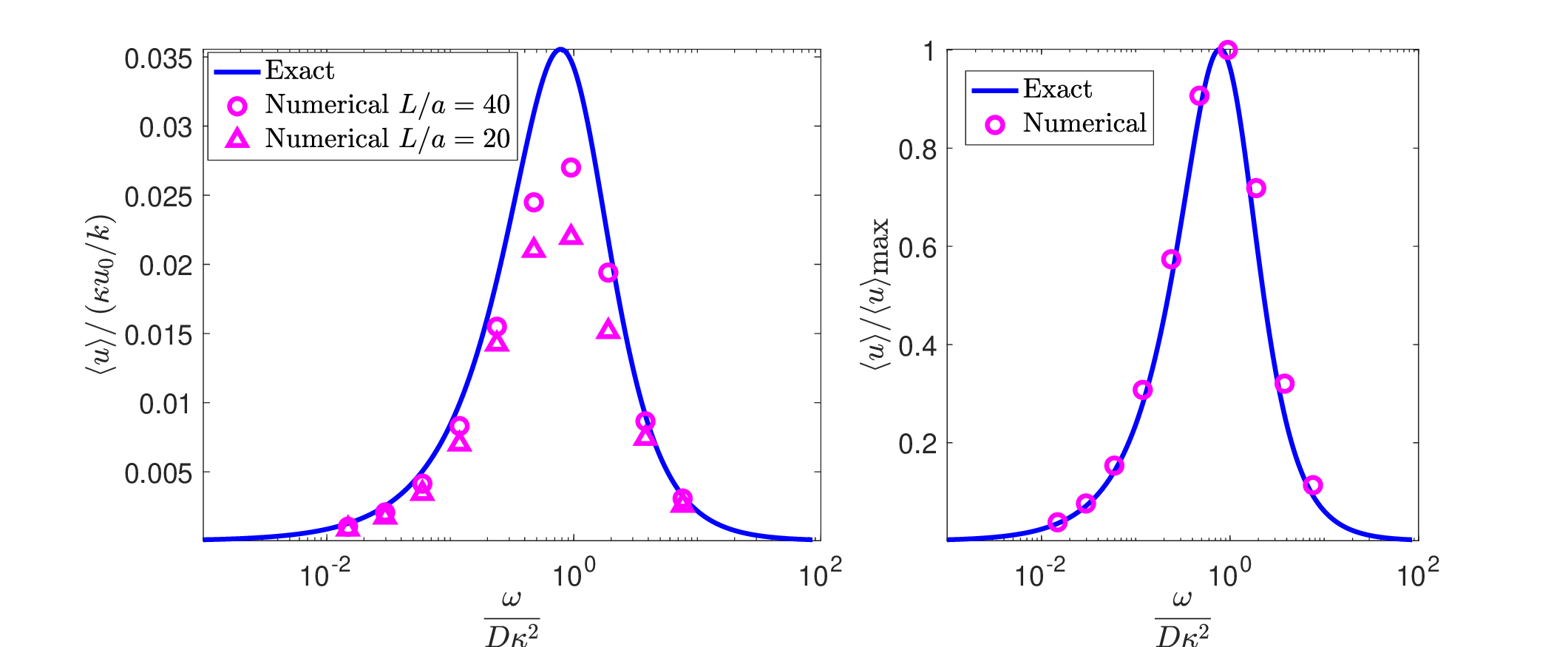}
\vspace{-0pt}
\end{center}
\caption{Left panel: Comparison of the exact
zero mode velocity \rr{uavcylinder} (continuous curve, calculated with a cylinder of infinite length) with two of its numerical counterparts (circles \& triangles) obtained with the finite-element package Comsol 
in the geometry displayed in figure \ref{cylinder}
by employing the parameter values 
\rr{paramN} (with cylinder aspect ratio $L/a=20$ and $L/a=40$, respectively). In general, the agreement between the exact and the numerical results is very good near the tails. The difference between numerical and exact solutions is attributed to the flow not being fully developed when the aspect ratio $L/a$ is relatively small. As the aspect ratio $L/a$ increases however, the numerical solution converges to the exact as seen by comparing the circle with the triangle data points.
%The dashed line denotes the single-parameter fit \rr{EM1982c} for the value $\beta = 1.47$.  
%If the value of $k$ becomes sufficiently large, say $k>10^5 \textrm{ m}^{-1}$, the amplitude of the curves
%decreases. 
Right panel: Following \citet[Fig. 10(b)]{Cahill2004}, we display the same data points/continuous curve as in the left panel but now dividing the velocities by their maximum value. 
In general, the agreement between exact and numerical results is excellent. 
\label{uomegacylinderN}  }
\vspace{-0pt}
\end{figure}

To obtain the zero mode we average the time-dependent velocity field 
over the cylinder cross-section, the cylinder length $L$ (cf. figure \ref{cylinder}) and over $N$ periods of oscillation of the applied field, where $N$ is an integer. The details are described in the Supplementary Materials addendum.

The circle and triangle symbols in the left panel of figure \ref{uomegacylinderN} denote the velocity obtained numerically with the
configuration of figure \ref{cylinder} and parameters \rr{paramN}.
% averaged over the cylinder cross-section, length $L$ and also averaged over $N$ periods of oscillation
%of the wall charge distribution.
In general, there is excellent agreement with the exact solution \rr{uavcylinder} 
(continuous curve) at the tails. Away from the latter, the 
numerical and exact results agree well in an order-of-magnitude basis and share the same general trend. 
%We include a good single-parameter fit for the numerical results given by the dashed curve in Fig. \ref{uomegacylinder}, 
%having the functional form \rr{EM1982c} and the value of the fit parameter is $\beta = 1.47$. 

The difference seen between
the averaged zero mode velocity obtained numerically here and the exact model, as seen in the left panel of figure \ref{uomegacylinderN},
can mostly be attributed to the requirement of finite length geometry
for the numerical configuration. Indeed, we have verified that by increasing $L$ the numerical results approach their 
analytical counterparts as can be seen by comparison of the circle and triangle data points on the left panel of figure \ref{uomegacylinderN} with its exact counterparts. One explanation for this behavior is that small length $L$ of the capillary tube is mainly associated with a hydrodynamic entrance flow and only leaves a small lengthwise
domain over which the flow is fully developed, while the analytical result assumes that the flow is fully developed everywhere.
Fig. \ref{uomegacylinderN} is generated by employing the parameter values \rr{paramN} where the tube aspect ratio  (length-to-radius) is $L/a = 20$ and $L/a=40$. In the Supplementary Materials addendum we provide the corresponding data tables. 
%plots analogous to the left panel of figure \ref{uomegacylinderN} that show the effect of increasing the ratio $L/a$ and thus allowing a longer domain for the flow to become fully developed.
A full parametric analysis of traveling-wave electroosmosis in a finite-length capillary tube is beyond the scope of this paper. 

Finally, following \citet[Fig. 10(b)]{Cahill2004}, we compare
the exact and numerically-obtained zero mode velocities by scaling each one of the curves in the right panel of figure  \ref{uomegacylinderN} 
by their respective maximum. Now, there is an excellent agreement between the exact and the numerically obtained
zero mode velocities as displayed in the right panel of figure \ref{uomegacylinderN}.

\section{Discussion}
The right panels of the key figures \ref{uz_inf_E_phi2} and \ref{uav_selfsimilar_phi_channel}  show that 
scalings required for the Dirichlet problem to bring the velocity in self-similar form differ between the semi-infinite
space and the channel.  In addition the frequency dimensionless groups depend on the wavenumber $k$
of the excitation at the wall. 
%It thus seems that use of Dirichlet boundary conditions to describe effects
%and experiments in traveling wave electroosmosis might not be the best candidate to reach self-consistent
%and universal conclusions. 
On the other hand, 
the left panels of the same figures (and their accompanying scaling forms
\rr{uavssEinf} and \rr{uavssEchannel}, both obtained with Neumann boundary conditions for the Poisson equation) show that the excitation frequency $\omega$ 
is scaled by its Debye counterpart $D\kappa^2$, that is independent of the excitation wavelength and does not
depend on geometry considerations. 
This latter quality might make Neumann boundary conditions to be preferable to their Dirichlet counterparts in modeling
traveling-wave electroosmosis-related effects as was already pointed-out in the literature of electrokinetic energy conversion and its efficiency in 
nanofluidic channels by matching theory \citep{vanderHeyden2006} with experiment \citep{vanderHeyden2007}.

Along with our theory we provided the simple \emph{single-parameter} fits $\langle  u \rangle_{E\&M}$ (see \rr{EM1982b} and \rr{EM1982c})
which can be employed as a rapid test of consistency between our theory and future traveling wave electroosmosis
experiments. 
 
The Neumann velocities increase with decreasing channel width $h$, reaching a plateau, cf. the continuous
curve in the right panel of Fig. \ref{uav_h_Ephi}. The Dirichlet velocities, on the other hand, after reaching a maximum, decay to zero as $h \rightarrow 0$, see the dashed lines in the right panel of Fig. \ref{uav_h_Ephi}
and the exact expression \rr{uavh0}. Likewise the velocity becomes more pronounced when the excitation wavelength 
increases, cf. Appendix \ref{sec: kscaling} and \citep[Fig. 6(f)]{Liu2018}.
These geometrical realizations may have consequences on how an experiment is designed. 

The velocity estimates we obtain, by employing Neumann boundary conditions for the Poisson equation, seem to be within the purview of experimental measurements, at least in an order-of-magnitude basis. We were currently unable to exactly match the available experimental results with our theory. We believe however that, this can be done in a future contribution after carefully selecting the boundary conditions the Poisson equation satisfies at a wall. 

The numerical simulations we performed with the finite-element package Comsol for the configuration displayed in 
figure \ref{cylinder} and described in section \ref{sec: numerical} and the Supplementary Materials addendum, demonstrate the consistency in order-of-magnitude and general trend between the exact zero mode velocity and its numerically-obtained counterpart, despite the 
finiteness of the configuration employed in the latter. Of course, the displayed numerical results can be calibrated by 
changing parameters and conditions (eg. by setting the electric field to be zero in the reservoirs or setting-up
periodic boundary conditions) and can be generalized by including the effect the advective terms in the Nernst-Planck equation have on the zero mode.

The formulas describing the observable of interest here, that is, the liquid velocity in \rr{uzchannel} and \rr{urcylinder}, are general and can thus
be adopted to different configurations than the ones employed here. This is attained by solving the 
commensurate electric problem with alternative boundary conditions and using the provided velocity formulas for the `slip'
\rr{uzpolarabs} and average velocities
\rr{uavchannel2} and \rr{uavcylinder2} in the main body of this article `as is'.

In general, the  zero mode velocity is attributed to the invariance of the equations of motion (see  for instance the vorticity equation \rr{psinl})
under $SO(2)$ rotations in the $x$-$z$ plane (in the channel case) or $z$-$r$ plane (in the cylindrical capillary case), 
and the presence of mean-field solutions that can be parameterized by the elements of the invariance group. 
%\be \label{psinl0}
%\partial_t \nabla^2 \psi = \eta \nabla^4 \psi  + \partial_x \rho \partial_z \phi - \partial_z\rho \partial_x \phi
%\ee
%has a nonlinear term that is invariant under $SO(2)$ rotations in the $x$-$z$ plane. 
Quadratic  fluctuations of these mean field solutions give rise 
to a longitudinal mode that costs finite energy to change their magnitude, and to a zero mode 
costing no energy in the long wavelength limit. See \citep[p. 214-219]{Negele1988} for a detailed discussion.

Finally, we note that the existence and regularity of solutions to the Nernst-Planck-Navier-Stokes system for 
Dirichlet boundary conditions in an infinite periodic channel or with time-dependence (and so similar to the 
configuration and conditions employed here, but not identical), was recently established by
\citet{Constantin2021b,Constantin2022}. 
\\\\
\noindent
\textbf{Acknowledgments}\\ 
This work was supported by the US National Science Foundation through the Northwestern University MRSEC grant number DMR-2308691.
\\\\
\textbf{Declaration of Interests}\\ The authors report no conflict of interest.

\begin{figure}
\vspace{5pt}
\begin{center}
\includegraphics[height=1.6in,width=5.6in]{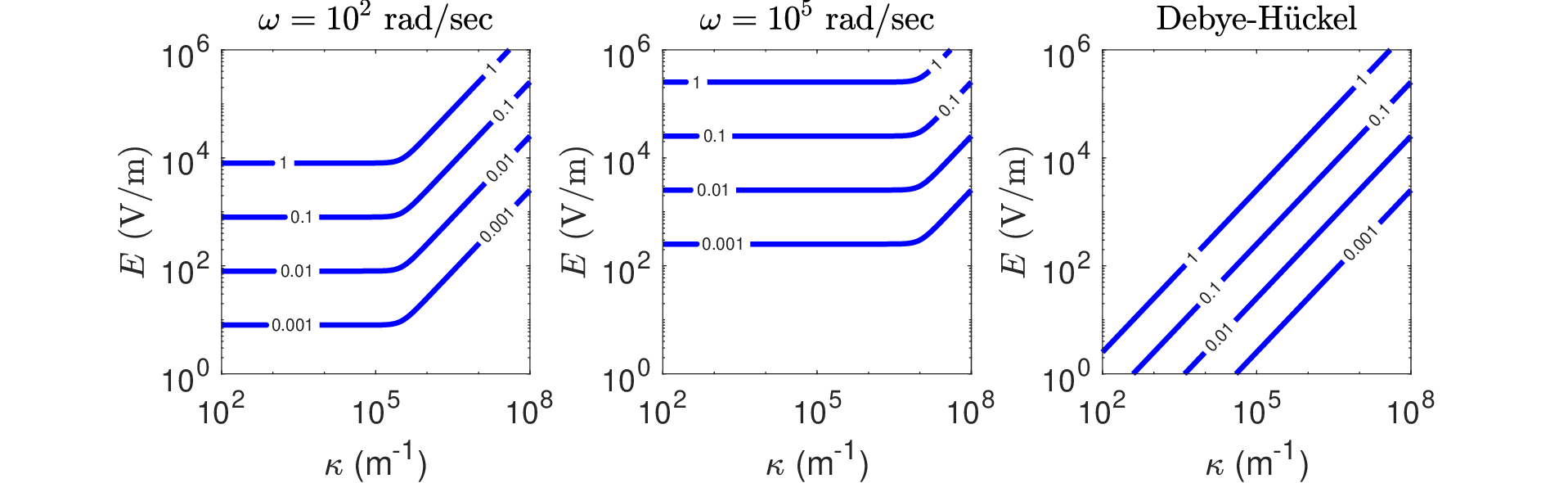}
\vspace{-0pt}
\end{center}
\caption{Validity of the charge perturbation approximation employed in this paper (contours of the function $f(\kappa, E) $ in \rr{cond0a} in the left two panels) and Debye-H\"uckel
approximation (contours of the function $\frac{eE}{\kappa k_BT }$ in  \rr{cond1} in the right-most panel). 
The plateau in the left two panels is due to the frequency renormalization of the Debye wavenumber $\kappa$
in the denominator of Eq. \rr{cond0a}. 
It is thus seen that
the range of validity of the approximation increases as the frequency $\omega$ (or the ratio $\omega/D$) increases.
Here $k=10^5\; \textrm{m}^{-1}$, and $D=10^{-9}\; \textrm{m}^{2}/\textrm{sec}$ at room temperature. 
\label{validityEkappa2}  }
\vspace{-0pt}
\end{figure}

\appendix
\section{The Debye-Falkenhagen approximation}
\subsection{\label{sec: validity}Validity of the Debye-Falkenhagen perturbation approximation}
We are unaware of a study in the literature examining the range of validity of the Debye-Falkenhagen approximation.
It is thus instructive to examine the conditions under which the reduction of the charge evolution equation \rr{rhot} and 
associated boundary condition was possible. The requirement is that $\rho \ll 2 ec_\infty$. 

For the semi-infinite space, Eq. \rr{rhoxzt} and \rr{RTheta} lead to the requirement
\be \label{cond0}
Re\left\{  \frac{\sigma_0 \kappa^2}{P}    \right\}<
\frac{\sigma_0 \kappa^2}{R(\kappa, k, \omega/D)} \equiv \frac{\epsilon E\kappa^2}{\left[ (\kappa^2+k^2)^2 + \left( \frac{\omega}{D}\right)^2 \right]^{\frac{1}{4}}} \ll 2 ec_\infty,
\ee
and this is analogous (but not identical) to the criterion reported by \citep[below their Fig. 5]{ehrlich1982} for the $k=0$ case. This condition then poses an upper bound
on the electric field amplitude in such a semi-infinite space. 

Eq. \rr{cond0} can be  reexpressed more clearly by eliminating $ec_\infty$  in favor of $\kappa$ through \rr{kappa}
\be \label{cond0a}
f(\kappa,  E) \equiv \frac{eE}{k_BT  \left[ (\kappa^2+k^2)^2 + \left( \frac{\omega}{D}\right)^2 \right]^{1/4} } \ll 1.
\ee
In retrospect, if we define an ``applied'' potential $\phi_a \equiv E/R$, Eq. \rr{cond0a} is nothing else than the statement
\be \label{condphi}
\frac{\phi_a}{\phi_T} \ll 1,
\ee
where $\phi_T \equiv  k_B T/e \sim 0.0253$ V, is the thermal potential. The bounds \rr{cond0a} and \rr{condphi} differ from the standard 
Debye-H\"uckel approximation (see Eq. \rr{cond1}) in that the wavenumber $\kappa$ in the latter is replaced here by the commensurate wavenumber $R(\kappa, k, \omega/D) = \left[ (\kappa^2+k^2)^2 + \left( \frac{\omega}{D}\right)^2 \right]^{1/4}$ defined in \rr{RTheta}. Thus, even for moderate frequencies $\omega\sim 10^2$ rad/sec, as shown in the left-most panel of figure \ref{validityEkappa2}, the approximation under consideration significantly improves compared to its Debye-H\"uckel counterpart (right-most panel of figure \ref{validityEkappa2}). This is the case because the wavenumber $R(\kappa, k, \omega/D)$ can become arbitrarily large when the frequency $\omega$ (present because
of the time-derivative in Eq. \rr{phit}), is moderate to high, even 
for small Debye lengths.

In the left two panels of figure \ref{validityEkappa2} we plot contours of the function $f(\kappa, E) $ in \rr{cond0a} for the constant values $f= 1, 0.1, 0.01$ and $0.001$  for $k=10^5\; \textrm{m}^{-1}$, and $D=10^{-9}\; \textrm{m}^{2}/\textrm{sec}$ at room temperature. It is thus seen that
the range of validity of the approximation increases as the frequency $\omega$ (or the ratio $\omega/D$) increases.

In the small Debye length limit Eq.  \rr{cond0} becomes
$\sigma_0 \kappa \ll 2 c_\infty$. Eliminating $c_\infty$ (or taking the large $\kappa$ limit in \rr{cond0a}) we obtain the following bound
\be \label{cond1}
\frac{eE}{\kappa k_BT } \ll 1.
\ee
This is analogous to the Debye-H\"{u}ckel approximation where $e\phi \ll k_BT$ \citep[p. 10.22]{Melcher1981}, if the 
characteristic length-scale of the system is the Debye length. 
In the right-most panel of 
figure \ref{validityEkappa2} we plot contours of the function $\frac{eE}{\kappa k_BT }$ at room temperature. Clearly the 
range of validity of the approximation leading to the bound \rr{cond0a} and displayed in the two leftmost panels of figure \ref{validityEkappa2}, is superior. 

In the bounded geometry of a channel of width $2h$, Eq. \rr{rhoxztchannel} leads to the requirement 
\be \label{cond0h}
Re\left\{  \frac{\sigma_0 \kappa^2 \cos Pz}{P\sin Ph}    \right\} \leq
\frac{\sigma_0 \kappa^2 \coth (P_2h)}{R(\kappa, k, \omega/D)} \equiv \frac{\epsilon E\kappa^2 \coth (P_2h)}{\left[ (\kappa^2+k^2)^2 + \left( \frac{\omega}{D}\right)^2 \right]^{\frac{1}{4}}} \ll 2e c_\infty,
\ee
where $P_{2}(\kappa, k, \omega/D) = \frac{1}{\sqrt{2}} \sqrt{R^2 + \kappa^{2}+k^{2}}$ was defined in 
\rr{P12} and $R$ in \rr{RTheta}. Eq. \rr{cond0h} differs from its semi-infinite space counterpart Eq. \rr{cond0} only by the factor $\coth (P_2h)$. 
For all practical purposes, this factor nearly equals $1$ and does not affect the approximation in a significant manner
when the frequency $\omega$ is non-vanishing.

\subsection{\label{sec: approximation}Comparison of the Debye-Falkenhagen with the Debye-H\"uckel approximation}
The vast majority of electroosmosis formulations employ the Debye-H\"uckel approximation which
amounts to setting the time-derivative in \rr{rhot} equal to zero. This gives a linear relation between $\rho$ and $\phi$, explicitly,
$\rho= -\epsilon \kappa^2 \phi$. On the other hand $\rho$ and $\phi$ in this paper are related through
\be
\rho = \epsilon \left(\frac{i\omega}{D} - \kappa^2\right) \left\{ \phi + \frac{\sigma_0}{\epsilon} 
\left[ \frac{\frac{i\omega}{D} }{\frac{i\omega}{D} - \kappa^2} \right]  \frac{e^{-kz}}{k} e^{i(kx -\omega t)}\right\}
\ee
cf. Eq. \rr{phixzt}. 
It clearly reduces to the form $\rho= -\epsilon \kappa^2 \phi$ in the limit $\omega \rightarrow 0$. 
Second, in the Debye-H\"uckel case, $\phi$ and $\rho \sim e^{-\kappa z}$ by solving either $\partial_z^2 \rho =\kappa^2 \rho $
or equivalently, $\partial_z^2 \phi =\kappa^2 \phi $. In contrast, in our case, $\rho \sim e^{-P_2 z} e^{i(P_1z +kx-\omega t)}$, 
where $P_{1}$ and $P_2$ were defined in \rr{P12}. 
Thus, in this paper the penetration depth is $1/P_2$. The  Debye-H\"uckel penetration depth $\kappa^{-1}$ has been renormalized by excitation wavenumber $k$ and frequency $\omega$, and species diffusion coefficient $D$. Plane waves propagate
in both the $x$ and the $z$ directions, the latter with a composite wavenumber $P_1$. 
%We briefly interject here to discuss the effect of the approximation that leads to the form \rr{rhot} for the 
%evolution of the charge distribution. Dropping the time-dependence and nonlinear terms, shows that the relation between
%charge and potential in the bulk of the liquid is 
%\be \label{rhophi1}
%\nabla^2 (\rho + \epsilon \kappa^2 \phi) = 0. 
%\ee
%On the other hand, the small potential (Debye-H\"{u}ckel) approximation employed in the vast majority of the literature leads to the following
%potential-charge relation in the bulk of the liquid
%\be \label{rhophi2}
%\rho + \epsilon \kappa^2 \phi =0,
%\ee
%where both \rr{rhophi1} and \rr{rhophi2} satisfy  the boundary condition $\rho + \epsilon \kappa^2 \phi =0$. 
%\rr{rhophi2} can be considered as a special case of \rr{rhophi1}. The two cases differ by a harmonic function which can have significant effects on whether the field can 
%satisfy certain boundary conditions. 

\section{Effect of the advection terms in the Nernst-Planck equation}
\subsection{\label{sec: linearization}Validity of droping the advection terms in the Nernst-Planck equation}
In the main body of this paper the advective term of \rr{rhot} was neglected on the basis of a small P\'eclet number
approximation. Here we justify the validity of this approach. 

We introduce new dimensionless variables
\be \label{ndimPe1}
X=kx, \quad Z = \kappa z, \quad \mathbf{V}= \frac{\mathbf{v}}{u_0}, \quad \tau = D\kappa^2 t,
\ee
where $u_0 = \frac{1}{2} \frac{\epsilon E^2}{\eta \kappa}$ is the velocity scale defined in 
\rr{u01}. 
Eq. \rr{rhot} becomes 
\be \label{rhotPe}
\partial_\tau \rho = \left[\left( \frac{k}{\kappa}\right)^2 \partial^2_X + \partial_Z^2 -1\right]\rho
- Pe \mathbf{V} \cdot \left( \frac{k}{\kappa} \partial_X, \partial_Z \right) \rho 
\ee
where
\be \label{Pe}
Pe = \frac{u_0}{\kappa D}.
\ee
This P\'eclet number scales quadratically with respect to the Debye length. 
Employing standard parameter values for water $\epsilon = 7\times 10^{-10}$ F/m, $\eta = 0.001$ kg/m sec, 
$\kappa = 10^{7}\textrm{ m}^{-1}$ (Debye length equal to hundreds of nanometers), $E =  10^5$ V/m 
and thus
$u_0\sim 0.35 \textrm{ mm/sec}$, we obtain the estimate
\be \label{estPe}
Pe \sim 0.035. 
\ee
When the Debye length is even smaller (say only tens of nanometers or the electric field smaller, say $E =  10^4$ V/m ), the P\'eclet number estimate \rr{estPe} is further reduced by two orders of magnitude.

%---------------------------------------------------------------------------
% figure(1) in nonlinear_advectionNDk.m
%---------------------------------------------------------------------------
\begin{figure}
\vspace{5pt}
\begin{center}
\hspace*{-1cm}
\includegraphics[height=2in,width=5.8in]{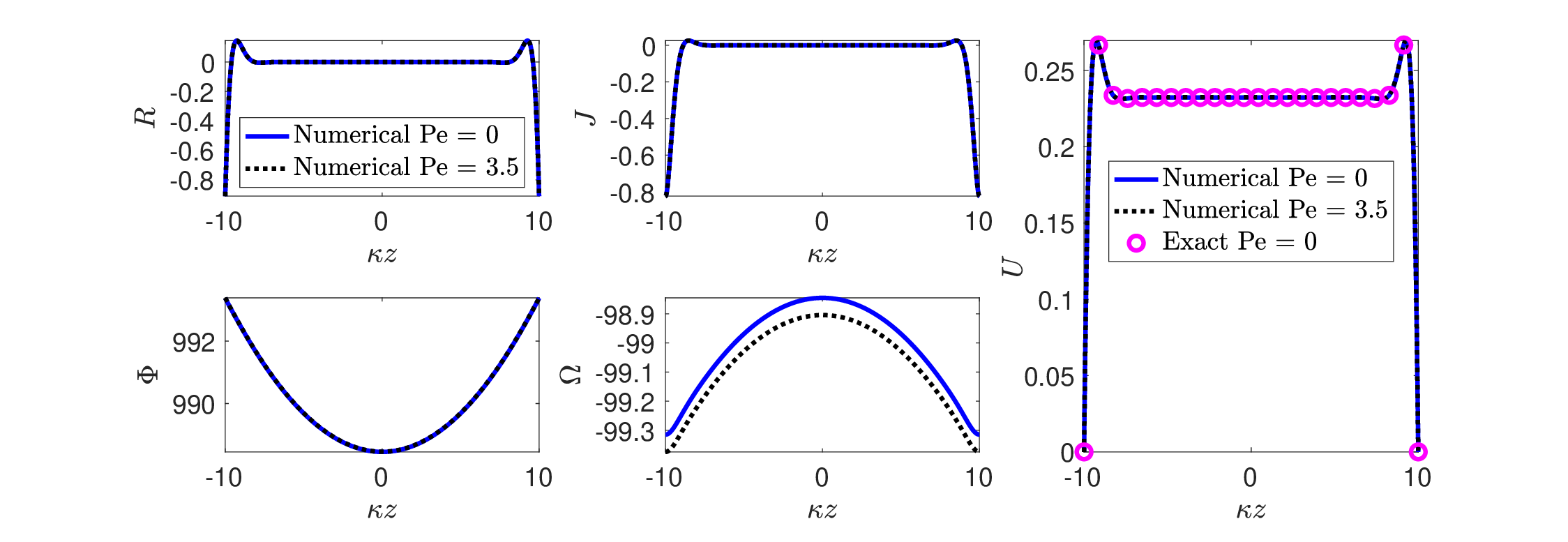}
\hspace{2mm}
\vspace{-0pt}
\end{center}
\caption{Determination of the effect of the P\'eclet number on the zero mode velocity in a channel.
Plot of the dimensionless fields $R,J, \Phi, \Omega$ and $U$ by solving
equations \rr{RJ} and \rr{PhiOmegaU} with boundary conditions \rr{systemBC}. The 
numerical solution for zero P\'eclet number (continuous curve) is nearly identical to the numerical solution
for an exaggerated value of 
$Pe=3.5$ (dotted curve). The circles denote the exact velocity \rr{uzchannel} (which was determined for zero P\'eclet number).  Here we have taken $\kappa =10^6 \textrm{ m}^{-1}$, 
$D = 10^{-9}\textrm{ m}^{2}/$sec, $k =10^4 \textrm{ m}^{-1}$, $h=10^{-5}$ m and $\omega = 10\times D\kappa^2 \textrm{ sec}^{-1}$.
\label{RJPhiOmegaU}  }
\vspace{-0pt}
\end{figure}

\subsection{\label{sec: Pe}The effect of P\'eclet number on the zero mode}
When the P\'eclet number is small, the fields can be expanded in a regular perturbation series with respect to this small number. In particular, the 
streamfunction will be of the form
\be \label{psiPe}
\psi = \psi_0(z) + \Pen \psi_1(z) e^{i\theta} + \Pen^2  \psi_2(z) e^{2i\theta} +...+c.c., 
\ee
where $\theta= kx-\omega t$, c.c. denotes complex conjugate terms and $\psi_0$ is the zero mode streamfunction ($u(z) = \partial_z \psi_0$). Such an expansion has been applied in the past to obtain amplitude equations of singularly perturbed nonlinear partial differential equations such as the Kuramoto-Sivashinsky and the 
Swift-Hohenberg equations cf. \citep{Kirkinis2014SH} and references therein as well as other linear and nonlinear problems
\citep{O'Malley2010a,O'Malley2011, Kirkinis2010a}. In this Appendix we will retain
only the leading order term in expansion \rr{psiPe} which will provide contributions to the advection term
in the Nernst-Planck equation \rr{rhotPe}. To be specific, we employ the channel geometry of 
section \ref{sec: channel}. 
%
%The equations of motion can be written as 
%\be
%\rho_{zz} + \frac{i\omega}{D} \rho - K^2 \rho = \frac{i k}{D} \rho u, \quad 
%\rho_{zz}^* - \frac{i\omega}{D} \rho^* - K^2 \rho^* =- \frac{i k}{D} \rho^* u, 
%\ee
Let
\be
\rho = R+iJ, \quad \rho^* = R-iJ, \quad \phi = \Phi +i \Omega, \quad \phi^* = \Phi - i \Omega.
\ee
for real fields $R, J, \Phi$ and $\Omega$. The dimensional form of the Poisson-Nernst-Planck-Stokes equations
for these fields becomes
\be \label{eq1a}
J_{zz} - K^2 J= -\frac{\omega}{D}R + \frac{k}{D}Ru, \quad R_{zz} - K^2 R =  \frac{\omega}{D}J - \frac{k}{D}Ju,
\ee
and
\be \label{eq1b}
\Phi_{zz} - k^2 \Phi = -\frac{R}{\epsilon}, \quad \Omega_{zz} - k^2 \Omega = -\frac{J}{\epsilon},\quad 
u_z = \frac{\epsilon k}{2\eta} \left (\Omega \Phi_z - \Phi \Omega_z \right),
\ee
where $K=\sqrt{\kappa^2 + k^2}$.
We also need nine boundary conditions
at $z=\mp h$
\be \label{dalPe}
R_z(\mp h) = \mp \sigma_0 \kappa^2, \quad J_z(\mp h)=0,\quad  \Phi_z(\mp h) = \pm \frac{\sigma_0}{\epsilon}, \quad \Omega_z(\mp h) = 0, \quad  
u(-h) = 0.
\ee

The above equations and boundary conditions are nondimensionalized by employing \rr{ndimPe1} and \rr{Pe}
and the fields
\be
(\hat{R},\hat{J}) = \frac{(R,J)}{2ec_\infty}, \quad (\hat{\Phi},\hat{\Omega}) = \frac{(\Phi,\Omega)\kappa}{E},
\quad
\hat{U} = \frac{u}{u_0}, \quad H = \kappa h
\ee
where $E$ is a characteristic scale for the applied electric field ($E\sim \sigma_0/\epsilon$), $u_0$ is the velocity scale defined in 
\rr{u01} and $h$ is the dimensional width of the channel. 

Equations \rr{eq1a}, \rr{eq1b} and their boundary conditions \rr{dalPe} become in dimensionless units
(dropping the hats)
\be \label{RJ}
R_{ZZ} - \left(\frac{K}{\kappa}\right)^2\!\!R =\frac{ \omega}{D\kappa^2} J-\frac{k}{\kappa} Pe JU, \quad J_{ZZ} - \left(\frac{K}{\kappa}\right)^2\!\!J = - \frac{ \omega}{D\kappa^2} R + \frac{k}{\kappa} Pe RU,
\ee
\be \label{PhiOmegaU}
\Phi_{ZZ} - \left( \frac{k}{\kappa} \right)^2\!\! \Phi =-\frac{\phi_T}{\phi_A} R, \quad 
\Omega_{ZZ} - \left( \frac{k}{\kappa} \right)^2\!\! \Omega =-\frac{\phi_T}{\phi_A} J, \quad 
U_Z = \frac{k}{\kappa} \left[ \Omega \Phi_Z - \Phi \Omega_Z \right]
\ee
and 
\be \label{systemBC}
R_Z(\mp H) =\pm \frac{\phi_A}{\phi_T}, \quad J_Z(\mp H) =0, \quad \Phi_Z(\mp H) = \mp 1, \quad \Omega_Z(\mp H) = 0, \quad U(-H) = 0,
\ee
where 
\be
\phi_A = \frac{E}{\kappa}, \quad \phi_T = \frac{k_BT}{e},
\ee
are the applied and thermal electric potentials, respectively and the P\'eclet number $\Pen$ was defined in \rr{Pe}.

Equations \rr{RJ} and \rr{PhiOmegaU} with boundary conditions \rr{systemBC} form a nonlinear first order system
of nine equations that is solved with a boundary value solver. In figure \ref{RJPhiOmegaU} we display the 
numerical solution of the system for zero P\'eclet number (continuous curve) and for an exaggerated value of 
$Pe=3.5$ (dotted curve). The circles denote the exact velocity \rr{uzchannel} (which was determined for zero P\'eclet number). 
The two numerical solutions of differing P\'eclet numbers agree very well. The conclusion is that the advection terms in the Nernst-Planck equation \rr{rhotPe} do not significantly alter the zero mode (when higher order harmonics can
be safely neglected).

\begin{figure}
\vspace{5pt}
\begin{center}
\includegraphics[height=1.2in,width=5.6in]{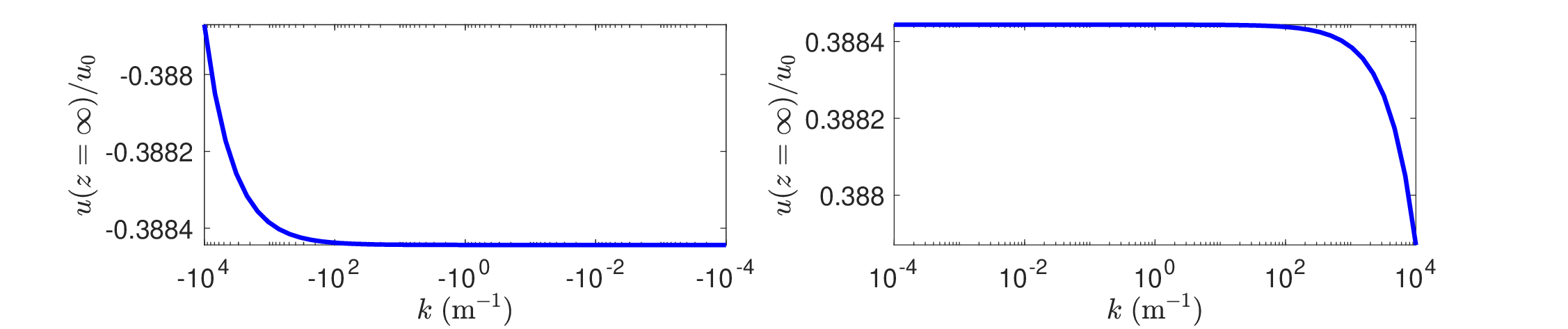}
\vspace{-0pt}
\end{center}
\caption{Negative and positive zero velocity branch \rr{uzpolarabs} displayed in the left and right panels, respectively, as a function of $k$. 
The velocity is discontinuous at $k=0$ which is \rr{EM1982}, cf.  \citep[Eq. (34)]{ehrlich1982}. Here we have taken $\kappa =10^7 \textrm{ m}^{-1}$ and
$D = 10^{-9}\textrm{ m}^{2}/$sec. 
\label{uz_inf_discont}  }
\vspace{-0pt}
\end{figure}

\section{\label{sec: E&M}The \citet{ehrlich1982} ``slip'' velocity}
The left and right panels of Fig. \ref{uz_inf_discont} display the plus/minus branch of the velocity \rr{uzpolarabs}, corresponding to positive/negative $k$, respectively.  
From fig. \ref{uz_inf_discont} it is apparent that the velocity at $k=0$ displays a discontinuity. This particular case has appeared before in the literature \citep{ehrlich1982}.  
Introducing the timescale (in the notation of \citet{ehrlich1982})
\be \label{Mnot1}
t_e \equiv  \frac{\epsilon k_BT}{2c_\infty e^2 D} = \frac{1}{D\kappa^2},
\ee
where $\kappa$ was defined in \rr{kappa}, and setting $k=0$
leads the amplitude and phase in \rr{RTheta} to obtain the form
$
R=\kappa  \left[1+\left( \omega t_e\right)^2\right]^\frac{1}{4}$ and $
\Theta = -\frac{1}{2}  \arctan \left( \omega t_e \right) +\frac{\pi}{2}$.
Substituting into \rr{uzpolarabs} and setting $k=0$ we obtain
\be \label{EM1982}
u(\infty) = \frac{\omega  t_e }{\left(\omega^{2} t_e^{2}+1\right)^{\frac{5}{4}}}\cos \! \left(\frac{\arctan \left(\omega  t_e \right)}{2}\right),
\ee
which is Eq. (34) of \citep{ehrlich1982} and $t_e$ was defined in \rr{Mnot1}. 
In the left panel of Fig. \ref{uz_inf_E_phi2} (continuous curve) we display \rr{EM1982}, which recovers (in logarithmic axes) Fig. 3 of \citep{ehrlich1982}.

%
%\section{\label{sec: timescales}Implicit and explicit timescales}
%There are many timescales involved in the electroosmosis problem as discussed by \citet{Bazant2004}, 
%for instance,
%\be \label{timescales}
%\tau_D = \frac{1}{D\kappa^2}, \quad \tau_c = \frac{h}{\kappa D}, \quad \tau_{k\kappa} = \frac{1}{Dk\kappa}, \quad \tau_k = \frac{1}{Dk^2}, \quad \tau = \frac{2\pi}{\omega}, \quad \tau_h = \frac{h^2}{D}
%\ee
%where $h$ is the width of the channel in section \ref{sec: channel} and $k$ and $\omega$ are the wavenumber and frequency of charge inhomogeneity at an insulated wall. In general $\tau_D \ll \tau_c\ll \tau$. In the solution of the \emph{initial value problem} for the Poisson-Nernst-Planck (PNP) equations, the first two timescales appear in the form of rapidly decaying exponentials $e^{-t/\tau_c}$ and $e^{-t/\tau_D}$ (see for instance 
%\citep[Eq. (19)]{Golovnev2012}) and thus they form what in the language of asymptotics is known as initial layers. 
%In the present paper we look for the long-time asymptotics (we do not solve the initial value problem) and thus
%the part of the solution involving these exponentials has already died-out. The charging time-scale $\tau_c$, however, appears 
%implicitly into the solutions of the PNP equations, for instance, the imaginary part of the combination $Ph$ appearing in equations \rr{rhoxztchannel} and 
%\rr{phixztchannel} can be written as 
%\be
%P_2 h = \frac{1}{\sqrt{2}} \sqrt{\sqrt{\left[(\kappa h)^{2}+(kh)^{2}\right]^{2}+ \left(\frac{\tau_c}{\tau} \right)^2(\kappa h)^{2}}+ (\kappa h)^{2}+ (kh)^{2}}.
%\ee
% 
% 
 \section{\label{sec: gravity}Relation of solution ansatz $\phi(z)e^{i(kx-\omega t)} $ to other literature}
 The form of the electric potential we have assumed to be true $\phi(z)e^{i(kx-\omega t)} $ in a space that is infinitely
 long in the $x-$direction, is of the same form employed to describe gravity waves in potential flow, that is, 
 when the velocity of the liquid can be written as the gradient of a scalar function $\phi$, which satisfies
 Laplace's equation with (nearly) Neumann (but homogeneous) boundary conditions \citep[\S12]{Landau1987}. For instance, the resulting velocity field of a liquid of depth $h$ obtains the form
 $$
 \mathbf{v}(x,z,t) = \frac{A\omega}{\sinh (kh)} \left( -\cosh(k(z+h))\sin (kx-\omega t),  \sinh(k(z+h))\cos (kx-\omega t)  \right)
 $$
 which essentially equals the electric field obtained from the channel case from Eq. \rr{phixztchannel} in the absence of an electrolyte.

\section{\label{sec: spacephi}Dirichlet problem: traveling wave electric potentials at a wall}
In the main body of this paper we employed Neumann boundary conditions that fix the normal electric field
on a charged wall. In the following discussion we will replace them with Dirichlet boundary conditions, that is, 
a fixed electric potential on the wall.

\subsection{\label{sec: semiinfinitephi}Semi-infinite space}
We consider the configuration displayed in Fig. \ref{space1wave} where now the traveling wave surface charges 
must be replaced by traveling wall electric potentials
on the wall $z=0$
bounding an $1:1$ electrolyte lying in the semi-infinite space $z>0$.
The boundary conditions satisfied by the charge $\rho$ and potential $\phi$ are
\be \label{bcinfphi}
\phi (x,z = 0) = \phi_0 e^{i (kx -\omega t)}, \quad \partial_z \rho(x,z = 0, t)  =   -\epsilon \kappa^2 \partial_z \phi(x,z = 0, t) , 
\ee
the latter being the 
 zero current condition $ \mathbf{J}\cdot \hat{\mathbf{z}} = - D\left[ \nabla \rho + \frac{e}{k_BT} s \nabla \phi  \right] \cdot \hat{\mathbf{z}}=0$ at the same wall leads to the requirement that $\partial_z \rho =   -\epsilon \kappa^2 \partial_z \phi$. 
These can be satisfied only after both fields are determined up to arbitrary constants
$
\rho(z) = A e^{iPz}, \quad \phi(z) = B e^{-kz} + \frac{A}{\epsilon(P^2  + k^2)} e^{iPz}. 
$
The charge distribution reads
\be \label{rhoxztphi}
\rho(x,z,t) = \frac{\epsilon\kappa^2k\phi_0 (P^2 + k^2)}{iP(P^2 + \kappa^2  + k^2) + k\kappa^2} e^{i(Pz +kx -\omega t)}, 
\ee
where $P$ is again expressed by Eq. \rr{kdelta2}. 
In the absence of potential modulation ($k=0$) the charge is zero everywhere. For nonzero $k$, the penetration depth is again determined by the 
inverse of the imaginary part of $P$ (taken to be positive).

Similarly, assuming $\phi = \phi(z) e^{i (kx -\omega t)}$, Eq. \rr{phit} reduces to $ \phi_{zz} -k^2 \phi =- \frac{\kappa^2k\phi_0 (P^2 + k^2)}{iP(P^2 + \kappa^2  + k^2) + k\kappa^2}  e^{iPz}, $ subject  to the 
boundary condition \rr{bcinfphi} 
and $\phi =0$ at infinity. 
Thus, the potential distribution reads
\be \label{phixztphi}
\phi(x,z,t) =\phi_0 \left( 1 - \frac{k\kappa^2}{iP(P^2 + \kappa^2  + k^2) + k\kappa^2} \right)  e^{-kz}e^{i(kx -\omega t)}
+\frac{1}{\epsilon (P^2 +k^2)} \rho, 
\ee
with $\rho$ given by \rr{rhoxztphi} and we assumed that $k >0$. 

%Thus, in the long wavelength limit $k\rightarrow 0$, 
%$\phi \rightarrow \phi_0$ and the electric field vanishes. However, this limit also makes the $z$ component of the
%electric field 
%\be  \label{Enormalphi}
%\mathbf{E}|_{z=0}\cdot \hat{\mathbf{z}} = \frac{i \phi_0 k \left(i R \sin \! \left(\Theta \right)+i k +R \cos \! \left(\Theta \right)\right) Re^{i(kx -\omega t)}}{i R^{2} \cos \! \left(\Theta \right)+i \cos \! \left(\Theta \right) \kappa^{2}-R^{2} \sin \! \left(\Theta \right)+\sin \! \left(\Theta \right) \kappa^{2}-R k}
%\ee
%at the wall to vanish, where $R$ and $\Theta$ were defined in \rr{RTheta}. Moreover, even in the nonzero $k$ limit, the normal component of the electric field at the wall $z=0$ in Eq. \rr{Enormalphi},
%is a complicated function of the traveling wave frequency $\omega$, wavenumber modulation $k$ and remaining material
%parameters that are implicit in the definition of the Debye wavenumber $\kappa$. Thus, knowledge of the amplitude $\phi_0$ of the traveling electric potential at a wall does not lead to an automatic 
%knowledge of the wall charge.  

\subsection{Channel case}
The boundary conditions satisfied by the charge $\rho$ and potential $\phi$ are
\be \label{bcchannelphi}
\phi (x,z =  \pm h) = \phi_0 e^{i (kx -\omega t)}, \quad \partial_z \rho(x,z = \pm h, t)  =   -\epsilon \kappa^2 \partial_z \phi(x,z = \pm h, t) , 
\ee
the charge distribution reads
\be \label{rhoxztchannelphi}
\rho(x,z,t) = \frac{\phi_0 \epsilon  \left(P^{2}+k^{2}\right) \sinh \! \left(k h \right) \kappa^{2} k \cos (Pz) e^{i(kx -\omega t)} }{P \sin \! \left(P h \right) \left(P^{2}+k^{2}+\kappa^{2}\right) \cosh \! \left(k h \right)+\sinh \! \left(k h \right) \kappa^{2} k \cos \! \left(P h \right)}
,  
\ee
where $P$ is the  complex wavenumber displayed in \rr{kdelta2}. Note that the denominator of expression \rr{rhoxztchannelphi} is composed of hyperbolic functions of large argument (since $P$ is complex) so, close
to the center of the channel ($z=0$) the charge is effectively zero, 
cf. \citep[Eq. (4)]{ajdari1995} for the corresponding case of a steady periodic
wall charge distribution. 
%This can be seen by averaging \rr{rhoxztchannel} over a two-dimensional cross-section of the channel,
%not just over the $z$ coordinate, cf. \citep[Eq. (4)]{Ajdari1995} for the corresponding case of a steady periodic
%wall charge distribution.  Had we chosen antisymmetric wall charges $\sigma^\pm =\pm \sigma_0 e^{i (kx -\omega t)}$, 
%the resulting bulk charge distribution replacing \rr{rhoxzt} would have been odd with respect to the origin $z=0$. Thus, the 
%vanishing of $\rho$ at $z=0$ cannot be employed as proof of electroneutrality in this problem, cf. \citep[Eq. (5)]{Ajdari1995}.

Similarly, assuming $\phi = \phi(z) e^{i (kx -\omega t)}$ and subject  to the 
boundary condition \rr{bcchannelphi}, the electric potential distribution reads
\be \label{phixztchannelphi}
\phi(x,z,t) = \frac{\sin \! \left(P h \right) P \left(P^{2}+k^{2}+\kappa^{2}\right) \phi_0\cosh (kz) e^{i(kx -\omega t)}}{P \sin \! \left(P h \right) \left(P^{2}+k^{2}+\kappa^{2}\right) \cosh \! \left(k h \right)+\sinh \! \left(k h \right) \kappa^{2} k \cos \! \left(P h \right)} 
+\frac{1}{\epsilon (P^2 +k^2)} \rho, 
\ee
with $\rho$ given by \rr{rhoxztchannelphi}. 

\begin{figure}
%   \centering
   \begin{subfigure}[t]{0.5\textwidth}
%        \centering
        \includegraphics[width=\linewidth]{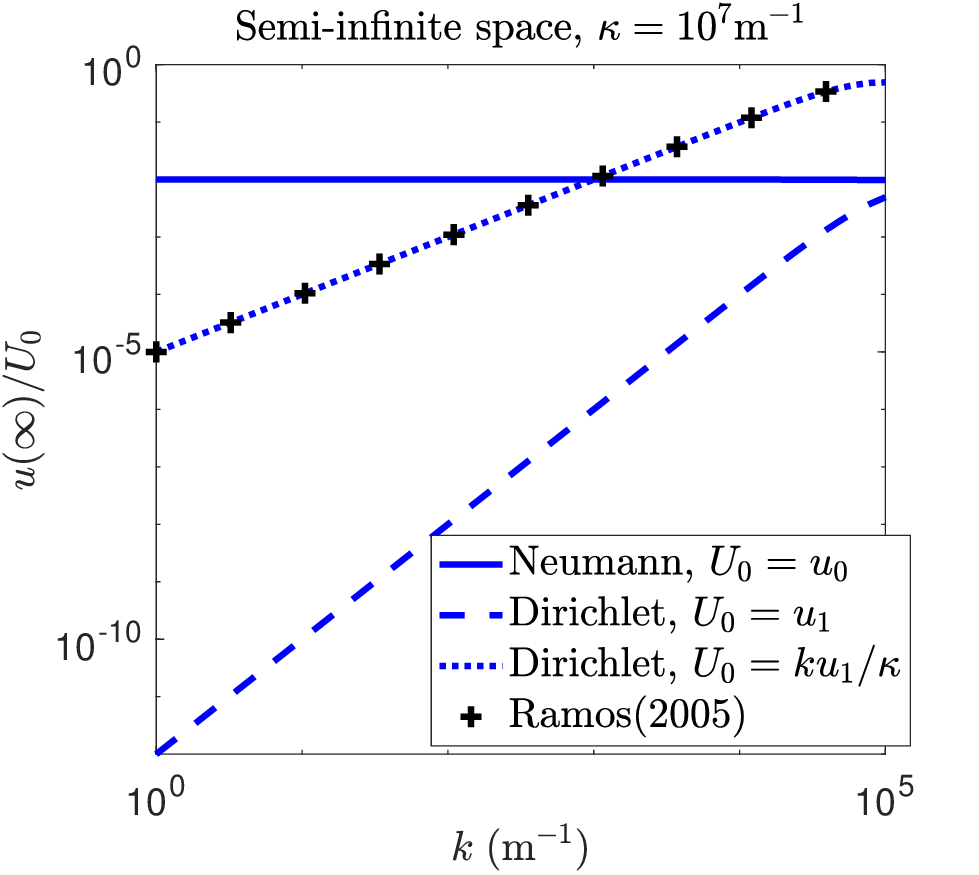} 
        \caption{} \label{uinfk}
    \end{subfigure}
%    \hfill
    \begin{subfigure}[t]{0.5\textwidth}
%       \centering
        \includegraphics[width=\linewidth]{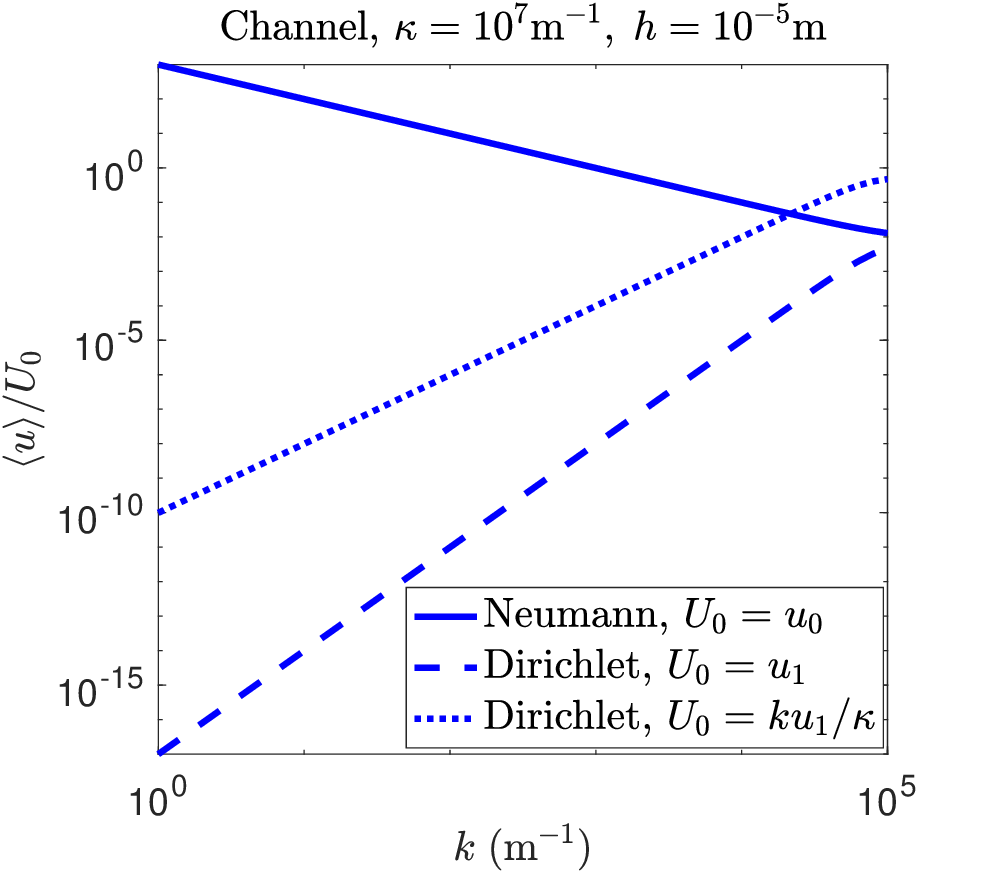} 
       \caption{} \label{uavk}
    \end{subfigure}
     %  \vspace{1cm}
%    \begin{subfigure}[t]{\textwidth}
%    \centering
%        \includegraphics[width=\linewidth]{example-image-c.pdf} 
%        \caption{Price regulation} \label{fig:timing3}
%    \end{subfigure}
%  \begin{subfigure}[t]{0.5\textwidth}
%%       \centering
%        \includegraphics[width=\linewidth]{odd_evanescent_Kelvinbei0} 
%%        \caption{Spatial distribution of the observables $v_p, \sigma^{\textrm{ch}}$ and $v^{\textrm{ch}}$.} \label{vpschvchN}
%    \end{subfigure}
    \caption{\label{u_k_loglog} Panel (a): Semi-infinite space zero mode velocity \rr{uzpolarabs} vs. the wall excitation wavenumber $k$, obtained with Neumann or Dirichlet boundary conditions (employing the potentials \rr{phixztchannel} and \rr{phixztchannelphi}, respectively) and scaled by either $u_0$ or $ku_1/\kappa$ (cf. \rr{u01}). Each line, from top to bottom, has slope $0, 1$ and $2$ respectively, determining the velocity $k-$ dependence as $k^{0}, k^1$ and $k^2$, respectively. 
The plus sign markers denote the theoretical velocity expression of \citet[Eq. (10)]{Ramos2005}, $\sim \frac{\Omega}{1+ \Omega^2}$, 
where $\Omega = \frac{\omega}{ Dk\kappa}$, which (nearly) coincides with the Dirichlet velocity when the scaling employed is $ku_1/\kappa$. The Dirichlet behavior is reminiscent of \citep[Fig. 6(a)]{Cahill2004}. 
Panel (b): Averaged over the channel width zero mode velocity \rr{uavchannel2} vs. the wall excitation wavenumber $k$, obtained with Neumann or Dirichlet boundary conditions and scalings as above.  Each line, from top to bottom, has slope $-1, 2$ and $3$ respectively, determining the velocity $k-$ dependence as $k^{-1}, k^2$ and $k^3$, respectively.  
The Neuman curve $k$ behavior is reminiscent of \citep[Fig. 6(f)]{Liu2018}, where the slip velocity increases
linearly with the excitation wavelength. 
In both panels $\omega = 10^3$ Hz. 
    } 
\end{figure}

\section{\label{sec: kscaling}The $k$ dependence of the velocities}
The preceding discussion, although informative, it does not provide an understanding of the magnitude of the 
effects, which we thus analyze in the present section. 
 
Figure \ref{uinfk} displays the $k$-dependence of the $u(\infty)$ zero mode velocity \rr{uzpolarabs} vs. the wall excitation wavenumber $k$, obtained with Neumann and with Dirichlet boundary conditions (employing the potentials \rr{phixztchannel} and \rr{phixztchannelphi}, respectively) and scaled by either $u_0$ or $ku_1/\kappa$  (cf. \rr{u01}).  Each line, from top to bottom, has slope $0, 1$ and $2$ respectively, determining the velocity $k-$ dependence as $k^{0}, k^1$ and $k^2$, respectively. It is thus seen that the Neumann curve is nearly constant
and essentially impervious to $k$ (as was also concluded in the left panel of Fig. \ref{uz_inf_E_phi2}). The theoretical 
expression for the velocity determined by \citet[Eq. (10)]{Ramos2005}, cf. Eq. \rr{Ramos1} below, is nearly
identical to the Dirichlet velocity when the scaling employed is $ku_1/\kappa$, in \rr{u01}. 

We repeat the above analysis but for the case of the channel in Figure \ref{uavk} which displays the $k$-dependence of the averaged over the channel width zero mode velocity \rr{uavchannel2} obtained with Neumann and with Dirichlet boundary conditions (employing the potentials \rr{phixztchannel} and \rr{phixztchannelphi}, respectively) and scaled by either $u_0$ or $ku_1/\kappa$ (cf. \rr{u01}).  Each line, from top to bottom, has slope $-1, 2$ and $3$ respectively, determining the velocity $k-$ dependence as $k^{-1}, k^2$ and $k^3$, respectively. The Neuman curve $k$ behavior is reminiscent of \citep[Fig. 6(f)]{Liu2018}, where the slip velocity increases
linearly with the excitation wavelength. 
The $k$ scaling behavior of the velocities of interest in this paper is summarized in table \ref{table: kEphi}. 

\begin{table}
  \begin{center}
\def~{\hphantom{0}}
  \begin{tabular}{c|c|c|}
  \hline
 &
Neumann &
Dirichlet \\
\hline
Semi-infinite space&$u(\infty)\sim$const.                    &$ u(\infty)\sim k^2$\\
Channel/Capillary&$\langle u\rangle \sim k^{-1}$ & $\langle u\rangle \sim k^3$\\
 Channel/Capillary&$\langle u\rangle \sim O(h^{0})$ & $\langle u\rangle \sim O(h^2)$
 \\
  \end{tabular}
\caption{Summary of scaling behavior of the (zero mode) velocity. Compare with Fig. \ref{u_k_loglog}. Angle brackets $\langle \cdot \rangle$ denote averaging over the width of the channel. We also
include the limits of the expressions as $h\rightarrow 0$ from Eq. \rr{uavh0} }
  \label{table: kEphi}
  \end{center}
\end{table}

 The difference in $k$ dependence of the field amplitude $\phi(z)$ for Neumann and Dirichlet boundary conditions, as this is displayed in table \ref{table: kEphi}, can be explained by 
resorting to a general principle known as Stokes' Rule \citep[p. 258]{Zauderer1989}: if $\phi$ satisfies Laplace's equation
with Neumann boundary conditions $\partial_z \phi = g(x)$, then the function $\chi(z) \equiv \partial_z \phi$ satisfies Laplace's equation
with Dirichlet boundary conditions of the form $\chi = g(x)$ on the same boundary. Thus, if $\chi\sim O(k^\alpha)$, 
then its integral with respect to $z$ is of $O(k^{\alpha-1})$, which is the behavior met here when solving the Poisson equation with Neumann and Dirichlet boundary conditions (to obtain this explicit behavior we tacitly assumed that $g(x)\sim e^{ikx}$ and thus both $\phi$ and $\chi$ vary harmonically in the $x$ direction).

\section{Simplification of the nonlinear electric force and torque in the momentum equation}
\subsection{\label{sec: nltorque}Simplification of the nonlinear torque in \rr{psinl}}
In this Appendix we reduce the fourth order vorticity equation \rr{psinl} to 
the first order \rr{upz}. 

Let
\be \label{rhoxphix}
2\rho = \rho(z) e^{i\theta} + \rho^*(z) e^{-i\theta}, \quad 2\phi = \phi(z) e^{i\theta} + \phi^*(z) e^{-i\theta}, 
\quad 2\rho_x = ik\left(\rho(z) e^{i\theta}  - \rho^*(z) e^{-i\theta}\right), 
\ee
etc., with a slight abuse of notation, where $\theta= kx-\omega t$. 
Thus, the nonlinear term in $\eta \partial_z^3u(z) = \rho_z \phi_x - \rho_x\phi_z$ becomes
\begin{align}
4(\rho_z \phi_x - \rho_x\phi_z)& = ik\left(\phi \rho_z^*-\phi^*\rho_z +\phi_z \rho^* - \rho \phi_z^* \right)\\
& = i \epsilon k \left(\phi_{zzz}\phi^* - \phi\phi_{zzz}^* + \phi_z^*\phi_{zz} - \phi_z\phi_{zz}^*\right) \\
& = i \epsilon k \left( \phi^*\phi_z - \phi \phi_z^*  \right)_{zz}
\end{align}
where we replaced $\rho$ by $\epsilon(k^2 \phi  - \phi_{zz})$.  
In summary, $4\eta \partial_z^3u(z) = i \epsilon k \left( \phi^*\phi_z - \phi \phi_z^*  \right)_{zz}$
or 
\be \label{nlt3}
u'(z) = \frac{i\epsilon k}{4\eta} \left( \phi^*\phi_z - \phi \phi_z^*  \right) + Az + B.
\ee
The two constants of integration drop out. In the semi-infinite case, where the velocity at the wall is zero and 
at $z=\infty$ finite, the expression on the right-hand side of
\rr{nlt3} vanishes (it is composed of decaying exponentials), so $A\equiv 0$, while performing an extra integration, 
leads to the vanishing of $B$ since the velocity can only be finite at $z=\infty$. Thus, $u(z) = \mathcal{G}(z) - \mathcal{G}(0)$
where $\mathcal{G}(z)$ is the integral of the first term on the right hand-side of \rr{nlt3}. Thus, $u(\infty) = - \mathcal{G}(0)$. 

Likewise, in the channel case, $A$ is the pressure  gradient in the $x$-direction 
of the momentum equation (which is zero), and $B$ vanishes on the basis of the antisymmetry of 
the  first term on the right hand-side of \rr{nlt3} with respect to $z$ (and since $\partial_z u|_{z=0 }=0$).

\subsection{\label{sec: ns}Equivalence of \rr{nlt3} to the time-dependent Stokes equations}
Equation \rr{nlt3} was derived by integrating twice the vorticity equation for the zero mode. 
Below we show how \rr{nlt3} can directly be derived from the Navier-Stokes equations. 

There is no pressure gradient in the $x$-direction, $\mathcal{P}_x=0$, otherwise it would give rise to a commensurate
pressure-driven flow. Thus, the $x$-component of the Navier-Stokes equation is simply
\be \label{nsrhophix}
\eta u_{zz} = \rho \phi_x
\ee
where the right-hand side implies that the observables are real. With the same notation as in \rr{rhoxphix}
and replacing $\rho$ by $\epsilon(k^2 \phi  - \phi_{zz})$ we obtain
\be
4\rho \phi_x = i\epsilon k^3 (\phi^2 - (\phi^*)^2 ) - i \epsilon k (\phi\phi_{zz} - \phi^*\phi_{zz} + \phi \phi^*_{zz} - \phi^*\phi_{zz}^*)
= i\epsilon k (\phi^*\phi_z - \phi\phi_z^*)_{z}
\ee
and in the last step we integrated over the period of oscillation. Thus, substituting into \rr{nsrhophix}
leads to \rr{nlt3}. 

It is also easy to show that the $z$-component of the Navier-Stokes equations leads to the pressure expression
\be
4\mathcal{P} = \epsilon( k^2 |\phi|^2 - |\phi_z|^2) 
\ee
after averaging over the period of oscillation. 
Thus the hydrodynamic pressure $\mathcal{P}$ only depends on the vertical coordinate $z$, leading $\mathcal{P}_x$ to vanish, 
as required by \rr{nsrhophix}. 
}
%\\\\
%\noindent
%\textbf{Acknowledgments}\\ 
%This work was supported by the US National Science Foundation through the Northwestern University MRSEC grant number DMR-2308691.
%\\\\
%\textbf{Declaration of Interests}\\ The authors report no conflict of interest.

\bibliographystyle{jfm}
\bibliography{/Users/jmason/Desktop/LEFTERIS/Bibliography/fluids,/Users/jmason/Desktop/LEFTERIS/Bibliography/disorder,/Users/jmason/Desktop/LEFTERIS/Bibliography/materialscience,/Users/jmason/Desktop/LEFTERIS/Bibliography/perturbations,/Users/jmason/Desktop/LEFTERIS/Bibliography/Compressible,/Users/jmason/Desktop/LEFTERIS/Bibliography/physiology}
%\bibliography{/Users/eks400/Documents/Bibliography/disorder,/Users/eks400/Documents/Bibliography/materialscience,/Users/eks400/Documents/Bibliography/perturbations,/Users/eks400/Documents/Bibliography/fluids,/Users/eks400/Documents/Bibliography/physiology,/Users/eks400/Documents/Bibliography/Compressible}

\end{document}